\documentclass[a4paper,11pt]{article}

\usepackage{jcappub} 

\usepackage[T1]{fontenc} 

\usepackage{graphicx}

\newcommand{\be}{\begin{equation}}
\newcommand{\ee}{\end{equation}}
\newcommand{\ba}{\begin{eqnarray}}
\newcommand{\ea}{\end{eqnarray}}

\newcommand{\bea}{\begin{eqnarray*}}
\newcommand{\eea}{\end{eqnarray*}}

\title{Lyman-$\alpha$ power spectrum as a probe of modified gravity}

\author[a]{Philippe Brax,}
\author[a]{Patrick Valageas}

\affiliation[a]{Institut de Physique Th\'eorique, Universit\'e Paris-Saclay, CEA, CNRS,\\
F-91191Gif sur Yvette, France}

\emailAdd{philippe.brax@ipht.fr}
\emailAdd{patrick.valageas@ipht.fr}

\abstract{
We investigate the impact of modified-gravity models on the Lyman-$\alpha$
power spectrum. Building a simple analytical modeling, based on a truncated
Zeldovich approximation, we estimate the intergalactic medium power spectrum
and the Lyman-$\alpha$ flux decrement power spectrum along the line of sight.
We recover the results of numerical simulations for $f(R)$-gravity models
and present new results for K-mouflage scenarios. We find that the shape
of the distortion due to the modified gravity depends on the model, through
the scale-dependence or not of their growth rate.
This is more clearly seen in the three-dimensional power spectrum
than in the one-dimensional power spectrum, where the line-of-sight integration
smoothes the deviation.
Whilst the Lyman-$\alpha$ power spectrum does not provide competitive
bounds for  $f(R)$ theories, it could provide useful constraints
for the K-mouflage models. Thus, the efficiency of the Lyman-$\alpha$ power spectrum
as a probe of modified-gravity scenarios depends on the type of  screening mechanism
and the related scale dependence it induces.
The prospect of a full recovery of the three-dimensional Lyman-$\alpha$ power spectrum from data would also lead to
stronger constraints and a better understanding of screening mechanisms.
}

\begin{document}
\maketitle
\flushbottom

\section{Introduction}
\label{sec:Introduction}

Soon after the discovery of the acceleration of the expansion of the Universe, the possibility
that modified gravity on cosmological scales could be at the origin of this late time acceleration
was  raised \cite{Clifton:2011jh}.
This led to a  renewed interest in massive gravity \cite{Hassan:2011zd,deRham:2010kj},
which gives a tantalising link between the amount of dark energy necessary
to generate the acceleration of the Universe and the graviton mass.
Another class of models involving  a new scalar field, akin to the scalar polarisation of massive
gravity, could also  lead to self-tuned  \cite{Babichev_2018} acceleration  using
the Horndeski \cite{Horndeski:1974wa,Deffayet:2011gz} (and beyond)
\cite{Gleyzes:2014dya} classification of trouble-free scalar theories in four dimensions.
In both cases, the number of theories passing the stringent constraint on the speed of gravitational
waves is extremely limited \cite{Creminelli:2017sry,Sakstein_2017,Akrami_2018}.
In this paper, we shall focus on models which do not tackle the acceleration issue per se.
They all contain a vacuum energy which is partially responsible for the late-time acceleration.
On the other hand, gravity is still modified and constrained on large cosmological scales.
We concentrate on models of the $f(R)$ type for which the astrophysical constraints imply
that deviations from the $\Lambda$-CDM template are limited \cite{Lombriser_2014}.
Nevertheless, for models of the K-mouflage type \cite{Brax:2014a}, where large clusters
are not screened \cite{Brax:2015aa} and see the effects of the scalar field with no suppression,
large-scale observables such as the Lyman-$\alpha$ forest are of relevance.

Such a type of modified gravity has mostly been studied on large linear scales
through its impact on the cosmic microwave background (CMB) anisotropies
(in particular the integrated Sachs-Wolfe effect), on weak lensing cosmic
shear statistics, on the matter power spectrum;
and on small nonlinear scales through the matter power spectrum,
the abundance of virialized halos and clusters of galaxies.
However, very few studies have considered the impact on the
Lyman-$\alpha$ forest statistics, see for instance \cite{Arnold2015}
for a work focusing on $f(R)$-gravity using numerical simulations.
In contrast, the Lyman-$\alpha$ power spectrum has become a useful probe
of dark matter candidates
\cite{Narayanan_2000,Viel:2013aa,Baur:2016aa,Irsic:2017aa,Garny_2018},
providing constraints on warm dark matter scenarios
that suppress the matter power spectrum on small scales because of their
non-negligible velocity dispersion.
More generally, the Lyman-$\alpha$ power spectrum is a probe
of the linear matter power spectrum and of the background cosmology
\cite{Croft_1998,Croft_2002,McDonald:2003aa,McDonald:2005ab}.
Therefore, it is worth investigating whether the Lyman-$\alpha$ power spectrum
could also provide useful constraints on modified gravity.
In particular, to pass Solar System tests of gravity, modified-gravity
scenarios must include screening mechanisms that ensure convergence
to General Relativity in high-density and small-scale environments.
Depending on the model, screening may even apply up to galactic
or cluster scales. In contrast, Lyman-$\alpha$ forest clouds, which
correspond to weakly nonlinear density fluctuations and scales greater than
virialized objects, should not be screened.
Moreover, being in the weakly nonlinear regime, Lyman-$\alpha$ statistics
could be better probes of the linear power spectrum than highly
nonlinear scales, which are difficult to predict and involve complex
effects, such as baryonic feedback or virialization processes, that can be
degenerate with modified-gravity impacts or partly damp the sensitivity
to linear modified-gravity effects.
In addition, modified-gravity models often induce a new scale dependence
for the matter power spectrum, which is typically enhanced as compared
with the LCDM prediction instead of being suppressed as in warm dark matter
scenarios, that could also modify the shape of the Lyman-$\alpha$ power spectrum.

In this paper, we present a first step to estimate the impact of modified-gravity
scenarios on the Lyman-$\alpha$ power spectrum.
We briefly describe in section~\ref{sec:Cosmological-scenarios} the $f(R)$ and
K-mouflage theories that we investigate in this paper. Then, we first consider
the probability distribution function (PDF) ${\cal P}(F)$ of the Lyman-$\alpha$ flux in
section~\ref{sec:PDF}.
We use a standard fluctuating Gunn-Peterson approximation and estimate the
impact of modified-gravity theories through their amplification of the growth of matter
fluctuations.
We find that the effect on the small-scale IGM physics remains modest.
Then, we build in section~\ref{sec:Lyman-flux-Pk} a simple model for the Lyman-$\alpha$
power spectrum.
It is based on a truncated Zeldovich approximation, which encodes the dependence
on cosmology, and bias parameters that we keep fixed to simplify the analysis.
We check that it provides a reasonable match to numerical simulations and
observations, in the case of the concordance LCDM cosmology.
Next, we apply this modeling to the modified-gravity scenarios.
We check that we recover the results of numerical simulations \cite{Arnold2015}
for $f(R)$ theories, and we present new results for K-mouflage models.
We discuss the impact of these modified-gravity scenarios and we conclude
in section~\ref{sec:conclusion}.

\section{Cosmological scenarios}
\label{sec:Cosmological-scenarios}

To investigate the possible impact of modified-gravity scenarios on Lyman-$\alpha$
forest statistics, we consider in this paper two well-studied examples,
the $f(R)$ theories and the K-mouflage models.
We will assume that the variations of the various parameters
(such as the IGM temperature) that define the mapping between the matter density
field and the Lyman-$\alpha$ statistics can be neglected.
Thus, we focus on the deviations that directly arise from the change of the
underlying matter power spectrum and the growth of structures.
In a more accurate computation, we can expect the parameters that describe
the small-scale hydrodynamics of the Lyman-$\alpha$ forest to show some dependence
on the underlying gravitational theory, as they must be related to the formation
of large-scale structures. However, in practice some of these parameters are marginalized
over as nuisance parameters, or are directly measured from observations, such
as the IGM temperature, and can be taken as external fixed inputs.
A detailed study of this issue would require dedicated numerical simulations.
In this paper, we adopt a simple approach where we assume the main dependence
on cosmology to arise from the direct dependence on the large-scale matter
distribution and we use the numerical simulations available for the standard
LCDM cosmology and the $f(R)$ theories to check that we recover the correct
order of magnitude of the various effects we investigate.
In this section, we briefly present the two modified-gravity scenarios that
we will consider.

\subsection{f(R) theories}
\label{sec:f(R)}

The $f(R)$-gravity theories are already very strongly constrained by cosmological
and astrophysical data, but they remain interesting as simple examples
of modified-gravity effects on the matter distribution.
Moreover, they are the only case where numerical simulations of the
Lyman-$\alpha$ power spectrum have been performed \cite{Arnold2015}.
This will allow us to check the validity of our analytical modeling.
More specifically, we concentrate on a class of $f(R)$ theories of the
Hu-Sawicki \cite{Hu:2007nk} type, where the action is given by
\be
S_{f(R)} = \frac{1}{16\pi {\cal G}_{\rm N}} \int d^4x \sqrt{-g} \, f(R)
\ee
with
\be
f(R)= R- 2\Lambda^2 - f_{R_0} \frac{R_0^2}{R} ,
\ee
where $\Lambda^2/8\pi {\cal G}_{\rm N}$ is the vacuum energy responsible
for the late time acceleration of the Universe.
Here $R_0$ is the Ricci curvature of the Universe now.
We consider the cases of such  $f(R)$ theories with
$f_{R_0}=-10^{-4}$, $-10^{-5}$ and $-10^{-6}$.
Let us recall that in the linear regime, the growth of density perturbations is enhanced
as compared to the LCDM cosmology.
This can be described in the evolution equation
of the matter density contrast by an effective Newton constant,
\be
{\cal G}_{\rm N} \to {\cal G}_{\rm N} [ 1 + \epsilon(k,a) ] , \;\;\;
\epsilon(k,a) = \frac{2\beta^2(a)}{1+m^2(a) a^2/k^2} ,
\label{eq:epsilon-def}
\ee
where the coupling strength $\beta$ and the mass $m$ of the new degree of freedom
are
\be
\beta = \frac{1}{\sqrt{6}} , \;\;\; m(a) = \frac{H_0}{c}
\sqrt{\frac{\Omega_{\rm m0}+4\Omega_{\Lambda 0}}{2 | f_{R_0} |}}
\left( \frac{\Omega_{\rm m0} a^{-3}+4\Omega_{\Lambda 0}}
{\Omega_{\rm m0}+4\Omega_{\Lambda 0}} \right)^{3/2} \sim 1 \, h / {\rm Mpc} .
\label{eq:beta-fR}
\ee
Thus, on large scales $k \ll a m$ we recover General Relativity and the LCDM cosmology,
as $\epsilon \simeq 0$, while on small scales $k \gg a m$ Newtonian gravity is amplified
by a factor $4/3$. On smaller nonlinear scales, nonlinear screening effects
come into play and ensure a convergence back to General Relativity.
Note that the coupling strength $\beta$ is fixed and of order unity,
and we recover the LCDM cosmology at low $| f_{R_0} |$ by pushing the mass
$m(a)$ to infinity.

\subsection{K-mouflage models}
\label{sec:K-mouflage}

The K-mouflage models are also scalar-tensor theories, but the additional
scalar field is massless and has a nonstandard kinetic term.
This provides another simple example of modified-gravity scenarios
that includes an alternative screening mechanism.
The $f(R)$ theories give rise to the chameleon screening mechanism
\cite{Khoury:2003aq,Khoury:2003rn},
where the additional scalar field obtains a higher mass in high-density
environments, which decreases the range of the fifth force and screens
compact objects.
In contrast, the K-mouflage screening relies on a derivative
screening \cite{Babichev:2009ee,Brax:2012jr,Brax:2014a},
due to the nonlinearity of the kinetic term, so that the fifth force is damped
in regions of large field gradients (or large Newtonian force), which gives
rise to a K-mouflage radius around compact objects within which one recovers
General Relativity.
On large linear scales, from the point of view of the matter distribution,
the main difference from the $f(R)$ theories is that the scalar field
being massless there is no scale dependence for the linear growing mode,
as in the standard LCDM cosmology, but only a time-dependent amplification.

The K-mouflage theories are characterised by the coupling strength of the scalar field to matter
$\beta_K$ and a Lagrangian kinetic function $K(\chi)$ that is nonlinear.
This function must behave like $-1$ when the kinetic energy of the scalar field is small
in the late-time Universe, to play the role of the cosmological constant.
Moreover, it must also satisfy the stringent tests of gravity in the Solar System,
like the perihelion advance of the moon \cite{Barreira2015}.
In this paper we take
\be
K(0)= -1 \;\;\; \mbox{and} \;\;\;
K'(\chi) = 1 + \frac{K_\star \chi^2}{\chi^2+\chi_\star^2} ,
\ee
where $\chi=- (\partial \phi)^2/2{\cal M}^4$,
and ${\cal M}^4$ is the dark energy scale.
The first derivative $K'(\chi)$ goes from $1$ at low $\chi$, as for the standard kinetic term,
to the large value $K_\star$ at high $\chi$, which gives rise to the screening mechanism
that damps the scalar field gradients and the fifth force in high-density environments.
We choose to illustrate our results with $\chi_\star=100$ and $K_\star =1000$.
We consider the case of a coupling constant $\beta_K=0.1$
(the $f(R)$ theories correspond to $\beta_f=1/\sqrt{6}$ as in Eq.(\ref{eq:beta-fR})),
to remain consistent with constraints from Big Bang Nucleosynthesis and
the Solar System.
In practice, $\bar{K}' \simeq 1$ for the background for $z \lesssim 6$,
so that the precise form of $K(\chi)$ does not play any role and our results are
set by the value of the coupling $\beta_K$. Indeed, in this model
clusters of galaxies are still in the unscreened linear regime of the scalar field
\cite{Brax:2015aa} and this is even more so for the Lyman-$\alpha$ forest clouds.
Then, in the linear regime the growth of matter density perturbations can again be
described by an effective Newton constant as in Eq.(\ref{eq:epsilon-def}),
but it is now scale independent, with
\be
\epsilon(a) = \frac{2\beta_K^2}{K'} \simeq 2 \beta_K^2 = 0.02 \;\;\; \mbox{for} \;\;\;
\beta_K = 0.1 .
\ee
Thus, the amplitude of the deviation from Newtonian gravity is smaller
than in $f(R)$ theories, as $\epsilon = 0.02$ instead of the maximum value $1/3$
reached in $f(R)$ theories. However, it holds on a greater range of scales,
nearly up to the horizon and down to a smaller screening radius.

In contrast with the $f(R)$ models, we cannot compare our results to numerical
simulations, which remain to be developed. However, on linear scales the
K-mouflage scenarios mostly differ from the LCDM cosmology by a time-dependent
effective Newton constant, without introducing new scales. Therefore,
at a qualitative level, we can expect their large-scale physics to remain even
closer to the LCDM cosmology than for the $f(R)$ theories, and our modeling
developed in the next sections should fare as well as for the $f(R)$ theories.

\section{Flux probability distribution function}
\label{sec:PDF}

\subsection{Modeling ${\cal P}(F)$}
\label{sec:P-F}

The first statistics we consider in this paper is the probability distribution
function ${\cal P}(F)$ of the Lyman-$\alpha$ flux $F$.
In the standard fluctuating Gunn-Peterson approximation \cite{Gunn_1965,Croft_1998},
the Lyman-$\alpha$ optical depth is proportional to the neutral hydrogen density
$n_{\rm HI}$. For a gas in photoionization equilibrium that is mostly ionized, this is proportional
to the square of the density multiplied by a temperature-dependent recombination factor, 
and for temperatures $T \sim 10^4$ K this gives
\be
\tau \propto \rho^2 T^{-0.7} \propto (1+\delta)^{\alpha} \;\;\;
\mbox{with} \;\;\;
\alpha = 2 - 0.7 (\gamma-1) ,
\label{eq:tau-delta}
\ee
where $(\gamma-1)$ is the exponent of the gas density-temperature relation.
This gives for the Lyman-$\alpha$ flux
\be
F = e^{-\tau} = e^{-A (1+\delta)^{\alpha}}  .
\label{eq:F-delta}
\ee
The factor $A$ depends on the HI photoionization rate, which is difficult
to measure independently. Following standard practice, we set $A$ by requiring
that the mean flux $\langle F \rangle$ matches the observational measurements.
The exponent $\gamma$ is typically $\gamma \simeq 1 - 1.6$ and goes to
$1.6$ at late times in the case of early reionization \cite{Hui:1997ab}.
Following observations and numerical simulations \cite{McQuinn2009},
we take $\gamma = 1.3$ and $T=2 \times 10^4$ K at $z=3$,
which gives $\alpha=1.79$.
To relate the Lyman-$\alpha$ flux to the matter distribution through
Eq.(\ref{eq:F-delta}), we need to speficy the smoothing scale
$x_s$ of the density contrast.
As in \cite{Gnedin:1998aa,Zaldarriaga2003a}, we write the comoving smoothing
wave number $k_s$ in terms of the Jeans wave number $k_{\rm J}$ as
\be
k_s = 2.2 \, k_{\rm J} \;\;\; \mbox{with} \;\;\;
k_{\rm J} = \frac{a}{c_s} \sqrt{4\pi{\cal G}_{\rm N} \bar\rho} , \;\;\;
c_s= \sqrt{ \frac{5 k_{\rm B} T}{3\mu m_p} } ,
\label{eq:ks}
\ee
where $a$ is the scale factor, $c_s$ the sound speed, and
$\mu \simeq 0.5$ the mean molecular weight.
The factor $2.2$ accounts for the fact that the Jeans length was smaller at earlier times,
which reduces the damping scale at a given redshift \cite{Gnedin:1998aa},
and we take $x_s = \pi/k_s$ for the smoothing radius.
Then, neglecting the scatter of the density-flux relation (\ref{eq:F-delta}),
we write the flux PDF as
\be
{\cal P}(F) = {\cal P}(\delta_s) \left| \frac{d\delta_s}{dF} \right| .
\label{eq:P-F-P-deltas}
\ee
We define the density probability distribution function ${\cal P}(\delta_s)$
through its cumulant generating function $\varphi_s(y)$ (which is also
given by its Laplace transform),
\be
{\cal P}(\delta_s) = \int_{-{\rm i}\infty}^{{\rm i}\infty}
\frac{dy}{2\pi{\rm i}\sigma_s^2} e^{[y \delta_s - \varphi_s(y)]/\sigma_s^2}
\;\;\; \mbox{with} \;\;\;
\varphi_s(y)  = - \sum_{n=2}^{\infty} \frac{(-y)^2}{n!}
\frac{\langle \delta_s^n\rangle_c}{\sigma_s^{2(n-1)}} ,
\label{eq:phi-def}
\ee
As described in appendix~\ref{app:generating-function}, we take for $\varphi_s(y)$
its early-time or low-variance limit, which is determined by the spherical
collapse dynamics, see also \cite{Valageas2002,BraxPV2012}.
Because the Lyman-$\alpha$ forest probes mildly nonlinear perturbations,
the generating function $\varphi_s(y)$ may have already somewhat departed from this
quasilinear result. However, this should still provide a more accurate shape than
the usual lognormal approximation and it allows us to estimate the impact of
modified gravity on the shape of the PDF through higher-order cumulants, beyond
the density variance.

On the other hand, to be consistent with the approach we use for the Lyman-$\alpha$
power spectrum, we compute the smoothed variance $\sigma_s^2$ from the
nonlinear truncated Zeldovich power spectrum defined in Eq.(\ref{eq:PL-trunc}) below.
This means that $\delta_s$ is the IGM density field associated with the IGM power spectrum
(\ref{eq:Pk-IGM}). It differs from the underlying nonlinear matter distribution
by the smoothing scale $x_s$ and by the use of the truncated Zeldovich approximation,
which provides a reasonable description of the large-scale weakly nonlinear matter
distribution while removing the irrelevant contributions from high-density virialized
halos that do not contribute to the Lyman-$\alpha$ forest.

\begin{figure}
\begin{center}
\epsfxsize=7.5 cm \epsfysize=5.8 cm {\epsfbox{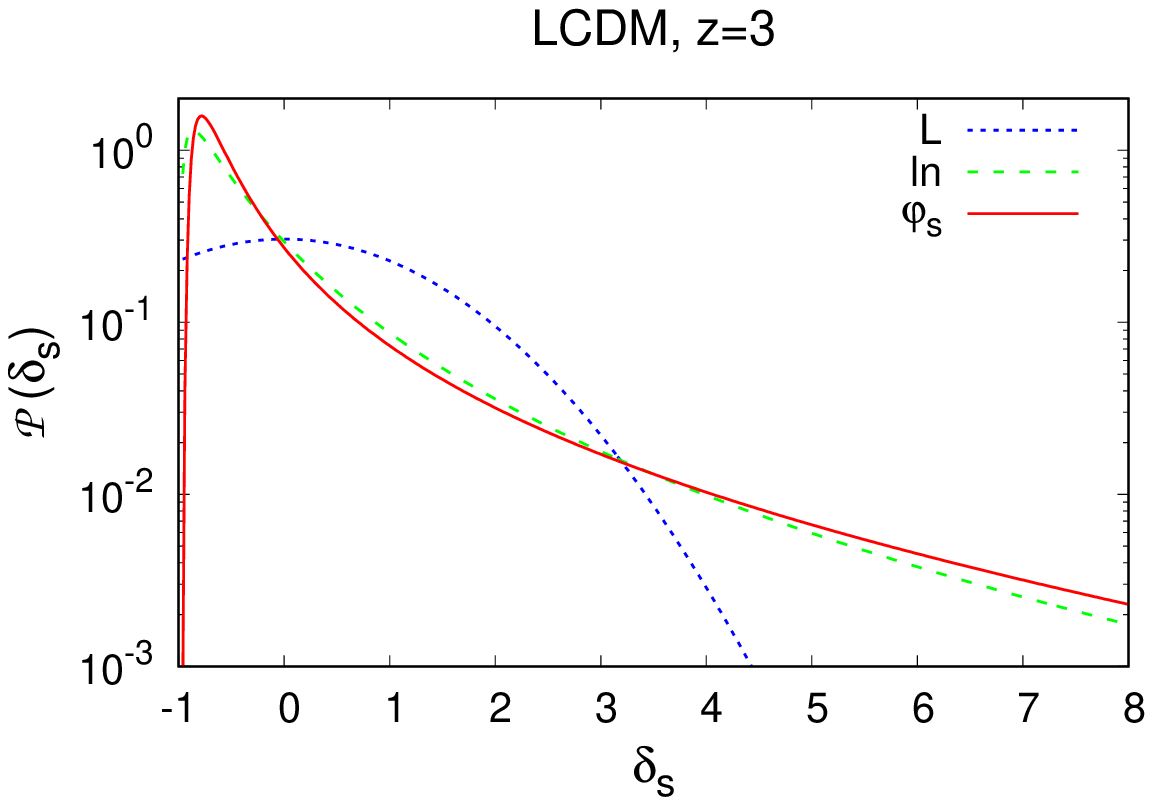}}
\epsfxsize=7.5 cm \epsfysize=5.8 cm {\epsfbox{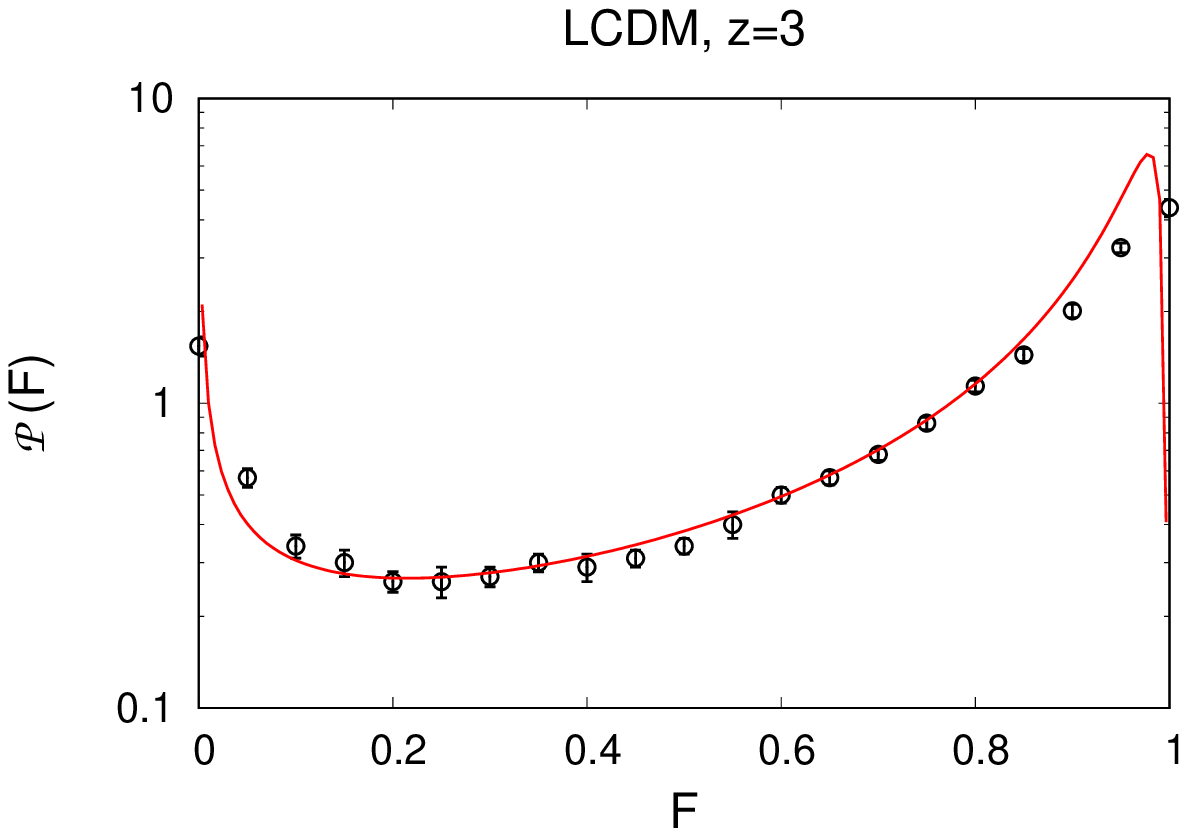}}
\end{center}
\caption{
{\it Left panel:} probability distribution function ${\cal P}(\delta_s)$ from
Eq.(\ref{eq:phi-def}) (red solid line labeled ``$\varphi_s$'').
We also display the Gaussian PDF from linear theory (blue dotted line ``L'')
and the lognormal approximation (green dashed line ``ln'').
{\it Right panel:} Probability distribution function ${\cal P}(F)$ from
Eq.(\ref{eq:P-F-P-deltas}).
The data points are the observational results of \cite{Calura:2012aa}.
}
\label{fig_Prho-z3}
\end{figure}

We compare in the left panel in Fig.~\ref{fig_Prho-z3} the density PDF (\ref{eq:phi-def})
with the Gaussian PDF from linear theory and the lognormal approximation.
We can see that on these mildly nonlinear scales, the density fluctuations of the IGM
are modest but the PDF already significantly deviates from the Gaussian,
with a peak at a slightly negative density contrast and an extended high-density tail.
As is well known, this shape is similar to the usual lognormal approximation.

Next, the mapping (\ref{eq:F-delta}) provides the flux PDF through
Eq.(\ref{eq:P-F-P-deltas}).
We can see in the right panel in Fig.~\ref{fig_Prho-z3} that this gives a reasonably
good agreement with the observations from \cite{Calura:2012aa}.
The coefficient $A$ in Eq.(\ref{eq:F-delta}) has been chosen so that the mean flux
matches the observed value of \cite{Calura:2012aa}, $\langle F \rangle = 0.72$,
which gives $A\simeq 0.5$.

\subsection{${\cal P}(\delta_s)$ for modified-gravity theories}
\label{sec:P-rho-modified-gravity}

\begin{figure}
\begin{center}
\epsfxsize=7.5 cm \epsfysize=5.8 cm {\epsfbox{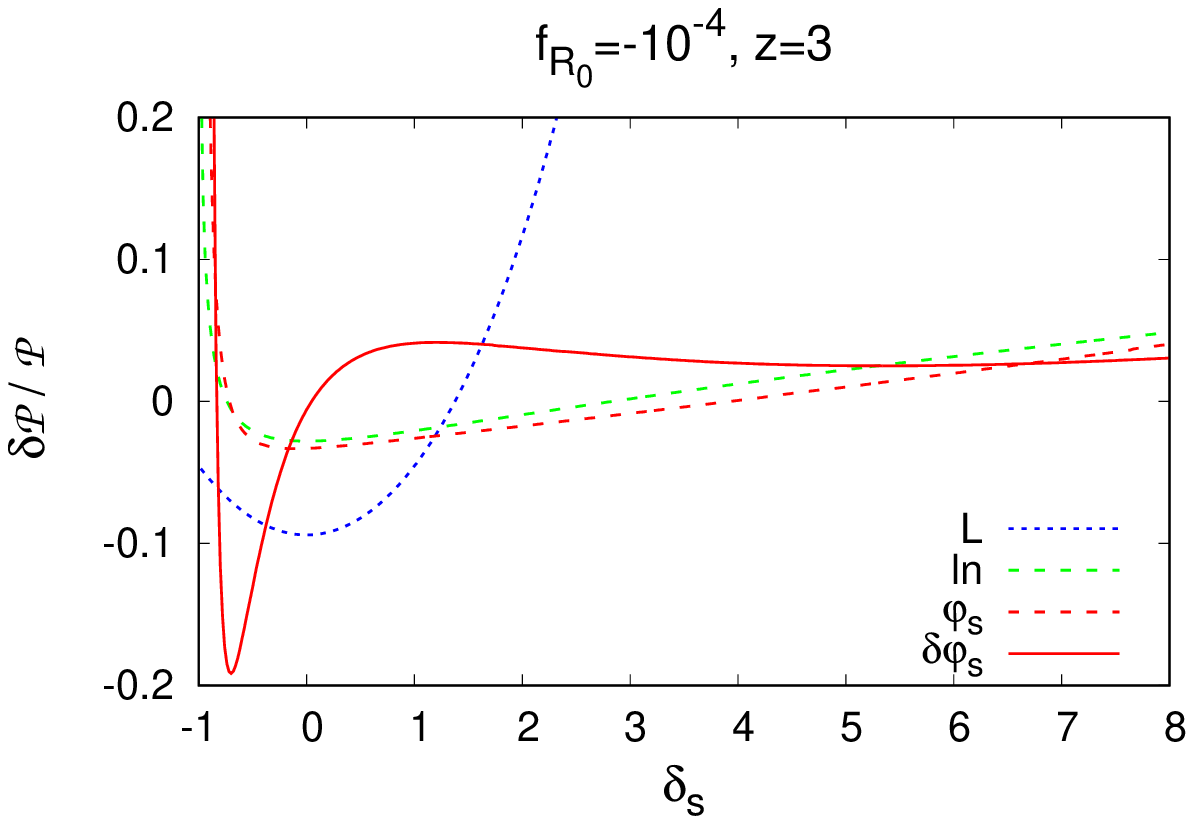}}
\epsfxsize=7.5 cm \epsfysize=5.8 cm {\epsfbox{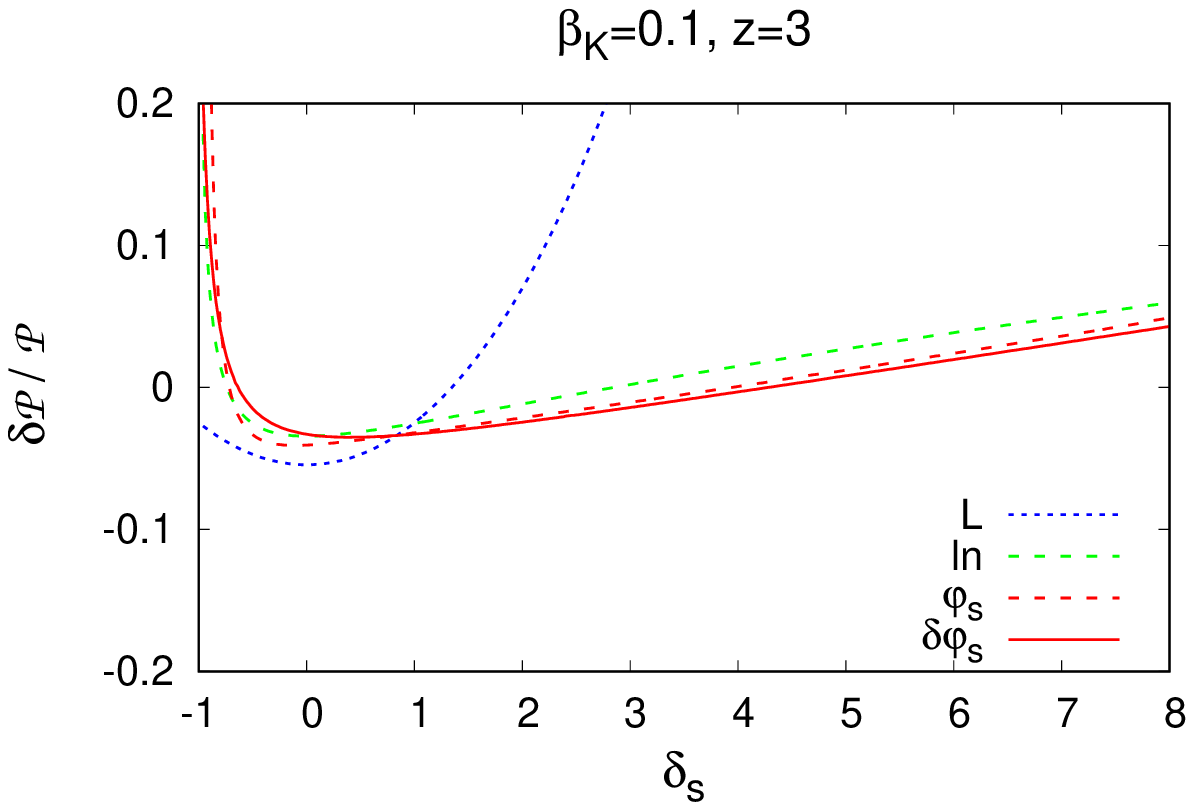}}
\end{center}
\caption{
{\it Left panel:} relative deviation of the PDF ${\cal P}(\delta_s)$ for the $f(R)$ models.
As in Fig.~\ref{fig_Prho-z3}, we display the Gaussian PDF from linear theory
(blue dotted line ``L''), the lognormal approximation (green dashed line ``ln''), and the
gravitational PDF (\ref{eq:phi-def}). For the latter, we consider the cases where we keep
the LCDM generating function $\varphi_s$ (red dashed line ``$\varphi_s$''),
or we use the new generating function determined by the modified-gravity spherical
collapse (red solid line ``$\delta\varphi_s$'').
{\it Right panel:} case of the K-mouflage model.
}
\label{fig_dPrho-z3}
\end{figure}

We show in Fig.~\ref{fig_dPrho-z3} the relative deviation of the density PDF
${\cal P}(\delta_s)$ from the LCDM prediction for the $f(R)$ theories and
the K-mouflage model.
We consider again the linear Gaussian PDF, the lognormal approximation, and the
expression (\ref{eq:phi-def}) that takes into account the nonlinear gravitational
dynamics. For the latter, we display the cases when we keep the LCDM function
$\varphi_s$ or use the modified-gravity function $\delta\varphi_s$
(by using the modified-gravity spherical collapse function ${\cal F}(\delta_{Lq})$
in Eq.(\ref{eq:phi-Legendre}) in appendix~\ref{app:generating-function}).
The modified-gravity scenarios studied in this paper amplify the growth of density
perturbations at low redshifts. This increases the variance $\sigma_s^2$ and
makes structure formation appear further advanced than in the LCDM cosmology.
This leads to stronger tails for the PDF ${\cal P}(\delta_s)$, as large fluctuations
are less rare, and hence to lower values of ${\cal P}(\delta_s)$ for moderate contrasts
$\delta_s \simeq 0$ as all PDFs are normalized to unity.
We recover this behavior in Fig.~\ref{fig_dPrho-z3}, with a positive deviation
$\delta{\cal P}$ at large negative and positive density contrasts.
Using either the lognormal approximation or the fixed LCDM generating function
$\varphi_s$ gives almost the same results for $\delta{\cal P}/{\cal P}$.
This agrees with Fig.~\ref{fig_Prho-z3}, which shows that the lognormal approximation
and Eq.(\ref{eq:phi-def}) lead to similar shapes, and the fact that in both cases
the deviation only arises from the change of the matter density variance.
For the $f(R)$ theories, the modification of the generating function $\varphi_s$,
that is, of the higher-order normalized cumulants
$\langle \delta_s^n\rangle_c/\sigma_s^{2(n-1)}$, has a significant impact.
It strongly amplifies the deviation from the LCDM result for moderate density
contrasts, $-0.8 \lesssim \delta_s \lesssim 3$.
In contrast, for the K-mouflage model the modification of the generating function
$\varphi_s$ is small and it has a negligible impact on $\delta{\cal P}$.
This expresses the fact that the K-mouflage gravitational dynamics is much closer
to the LCDM case than the $f(R)$ theories.
Indeed, as recalled in section~\ref{sec:K-mouflage}, on these scales the K-mouflage
scalar field $\phi$ is still in the linear regime (the nonlinear screening mechanism
only appears at galactic scales and overdensities). Then, the only modification
to gravitational processes is a slow time dependence of Newton's constant.
Clearly, this will not give rise to dramatic qualitative phenomena.
In contrast, for the $f(R)$ theories there is a new scale dependence, set by $m(a)$,
and the nonlinear regime of the modified gravity sector also becomes relevant faster.
Besides, the coupling strength $\beta$ is greater.
Therefore, it is not surprising that gravitational processes deviate more strongly from
the LCDM cosmology.

This can also be seen at the quantitative level in the coefficient $\nu_2$ obtained
from second-order perturbation theory, which would also enter the computation of bias
parameters. Thus, if we decompose the initial linear density perturbation into
a long wavelength mode $\delta_{Ll}$ and a short wavelength mode $\delta_{Ls}$,
$\delta_L=\delta_{Ll}+\delta_{Ls}$, higher-order perturbation theory
$\delta \simeq \sum_n \delta_L^n$ gives rise to mode couplings.
At second order in the density, and linear order in $\delta_{Ls}$, this gives
a contribution
\be
\delta_s = (1+ \nu_2 \delta_{Ll} ) \delta_{Ls} ,
\label{eq:nu2-def}
\ee
which describes how small-scale perturbations are enhanced by large-scale modes.
For the simple Einstein-de Sitter cosmology, this gives the well-known factor
$\nu_2=34/21$ that also corresponds to the angular average of the second-order
perturbation theory kernel $F_2({\bf k}_1,{\bf k}_2)$ \cite{Bernardeau2002}.
The expression (\ref{eq:nu2-def}) can also serve as a basis for the computation
of bias parameters \cite{Seljak_2012}.
On the other hand, on large scales the skewness of the density contrast,
which is the coefficient of the cubic term in the generating function $\varphi_s(y)$
and is defined by $S_3 = \langle \delta^3\rangle_c/\sigma^4$, is given by
$S_3 = 3 \nu_2 + d \ln \sigma^2/d\ln x$ \cite{Bernardeau2002}.
We give in Eq.(\ref{eq:nu2}) in appendix~\ref{app:second-order} the expression
of the angle-averaged coefficient $\nu_2(k_1,k_2)$ for general cosmologies.

\begin{table}
\centering
\begin{tabular}{|c|c|c|c|c|c|}
\hline
model &  LCDM & $f_{R_0}=-10^{-4}$ & $f_{R_0}=-10^{-5}$ & $f_{R_0}=-10^{-6}$ & $\beta_K=0.1$ \\
\hline
$\nu_2(k_s,1 h{\rm Mpc}^{-1})$ & 1.62 & 1.65 & 1.65 & 1.64 & 1.62 \\
$\nu_2(k_s,k_s)$ & 1.62 & 1.62 & 1.63 & 1.63 & 1.62 \\
\hline
\end{tabular}
\caption{\label{tab:nu2} Coefficient $\nu_2(k_1,k_2)$ for the LCDM, $f(R)$ and
K-mouflage cosmologies, at $z=3$. We show the cases
$(k_1,k_2)=(k_s,1 h{\rm Mpc}^{-1})$ (first row) and $k_1=k_2=k_s$ (second row).
}
\end{table}

We show the values of $\nu_2$ in Table~\ref{tab:nu2}.
For cosmologies with a scale dependence such as the $f(R)$ theories, it depends
on wavenumbers. We show in the first row the case of the pair $(k_s,1 h{\rm Mpc}^{-1})$,
where $k_s$ is the smoothing scale of Eq.(\ref{eq:ks}).
This describes the amplification of short-scale modes $k_s$ by long-scale modes
$k_l = 1 h{\rm Mpc}^{-1}$ as in Eq.(\ref{eq:nu2-def}).
The second row corresponds to the single wavenumber $k_s$, which is relevant for the
skewness of the density contrast at the single scale $k_s$.
Because the modified-gravity theories we study here amplify and speed up the linear
perturbations, they increase $\nu_2$ and the skewness of the PDF ${\cal P}(\delta_s)$.
For the scale-independent LCDM and K-mouflage cosmologies, the two rows are equal,
while for the scale-dependent $f(R)$ theories the value is higher and further from the
LCDM result in the first row, associated to different scales.
This is because these $f(R)$ theories introduce a new dependence on the ratio
$k_1/k_2$.
In agreement with Fig.~\ref{fig_dPrho-z3}, the deviation of $\nu_2$ from the LCDM value
is greater for the $f(R)$ theories than for the K-mouflage model, where it is below the
percent level.
This is due to the fact that the K-mouflage scalar field $\phi$ is still in the linear unscreened
regime and that in this model $\delta\phi$ is an odd functional of $\delta$, which implies
$\gamma^s_{2;11}=0$ for the new vertex that generically appears in Eq.(\ref{eq:nu2})
\cite{Brax:2014b}.
This explains the greater impact of the deviation of the cumulant generating function
$\varphi_s$ for the $f(R)$ theories.

\subsection{${\cal P}(F)$ for modified-gravity theories}
\label{sec:P-F-modified-gravity}

\begin{figure}
\begin{center}
\epsfxsize=7.5 cm \epsfysize=5.8 cm {\epsfbox{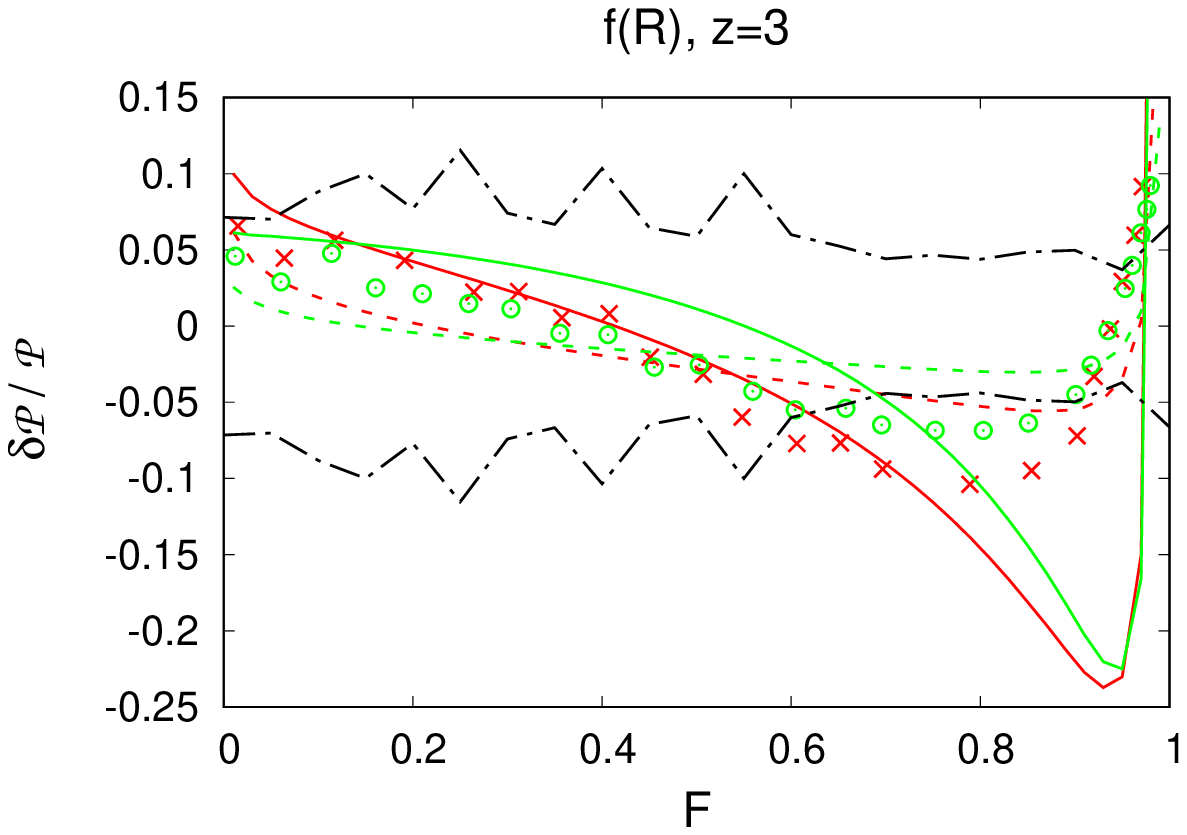}}
\epsfxsize=7.5 cm \epsfysize=5.8 cm {\epsfbox{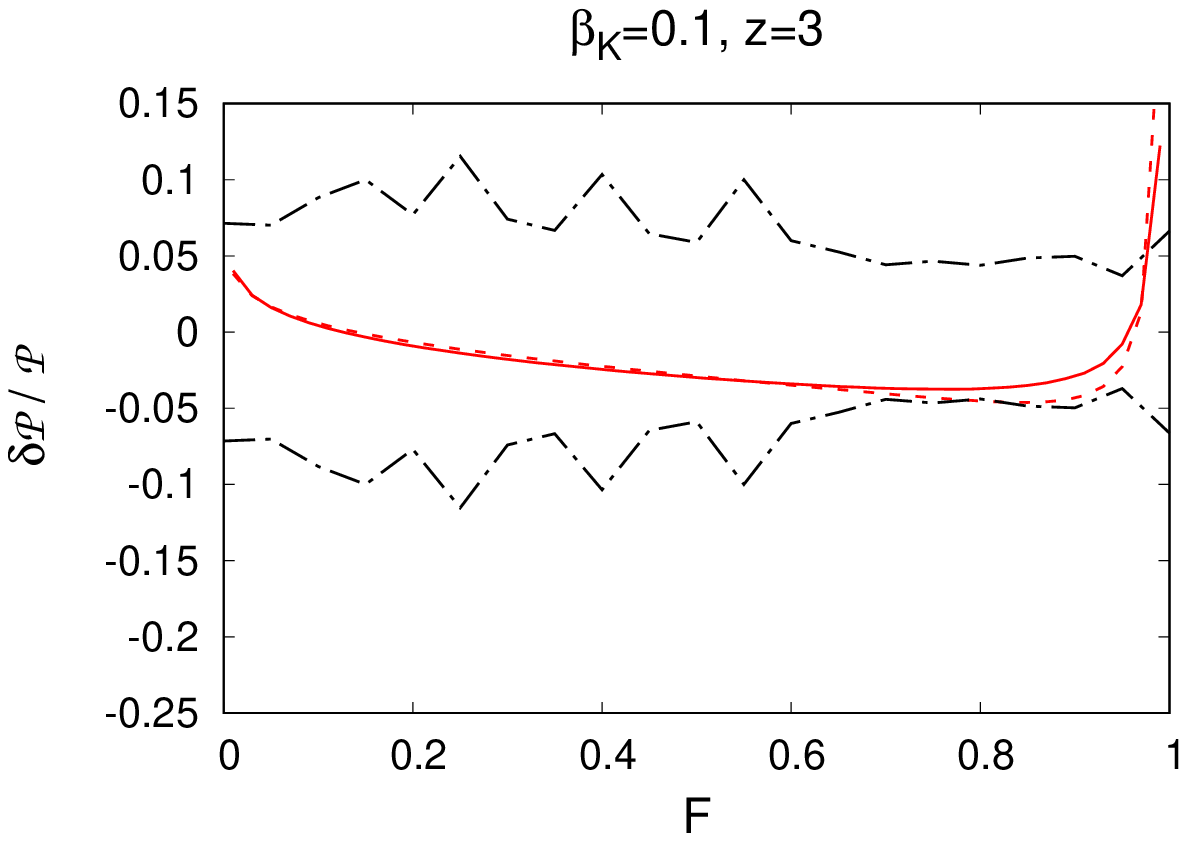}}
\end{center}
\caption{
{\it Left panel:} relative deviation of the PDF ${\cal P}(F)$ for the $f(R)$ models
$f_{R_0}=-10^{-4}$ (red lines) and $f_{R_0}=-10^{-5}$ (green lines).
The points are the numerical simulations of \cite{Arnold2015} for
$f_{R_0}=-10^{-4}$ (red crosses) and $f_{R_0}=-10^{-5}$ (green circles).
As in Fig.~\ref{fig_dPrho-z3}, the dashed lines neglect the dependence on cosmology
of the cumulant generating function while the solid lines use the modified generating function.
The symmetric upper and lower black dot-dashed lines are the $\pm 1\sigma$
relative errors of the observational results of \cite{Calura:2012aa}.
{\it Right panel:} relative deviation of the PDF ${\cal P}(F)$ for the K-mouflage model
(red lines), with the $\pm 1\sigma$ relative observational errors of \cite{Calura:2012aa}.
}
\label{fig_dPF-z3}
\end{figure}

We show in Fig.~\ref{fig_dPF-z3} the relative deviation of ${\cal P}(F)$ for the
$f(R)$ theories and the K-mouflage model.
As in the numerical simulations \cite{Arnold2015}, in all cases we set the coefficient $A$
in Eq.(\ref{eq:F-delta}) so that the mean flux matches the observed value of
\cite{Calura:2012aa}, $\langle F \rangle = 0.72$.
As for the density PDF shown in Fig.~\ref{fig_dPrho-z3}, the amplification of structure
formation in the modified-gravity scenarios leads to stronger tails for ${\cal P}(F)$,
and therefore to a lower amplitude of the PDF at the moderate values around
$\langle F \rangle$.

\begin{table}
\centering
\begin{tabular}{|c|c|c|c|c|c|}
\hline
model &  LCDM & $f_{R_0}=-10^{-4}$ & $f_{R_0}=-10^{-5}$ & $f_{R_0}=-10^{-6}$ & $\beta_K=0.1$ \\
\hline
$\langle \delta_F^2 \rangle$ & 0.199 & 0.215 & 0.212 & 0.211 & 0.202 \\
relative deviation & 0 & $8 \%$ & $7\%$ & $6\%$ & $1\%$ \\
\hline
\end{tabular}
\caption{\label{tab:F2} {\it First row:} variance $\langle \delta_F^2\rangle$ of the Lyman-$\alpha$ flux
for the LCDM, $f(R)$ and K-mouflage cosmologies, at $z=3$.
{\it Second row:} relative deviation of $\langle \delta_F^2\rangle$ from the LCDM prediction.
}
\end{table}

For the $f(R)$ theories we roughly recover the order of magnitude and the shape of the deviation
found in the numerical simulations \cite{Arnold2015}.
However, the agreement is not very good and the simulation results fall between the
two predictions obtained from Eq.(\ref{eq:phi-def}), where we either neglect or include the
impact of modified gravity on the generating function $\varphi_s$.
This suggests that the exact result is between these two approximations.
Indeed, the smoothing scale $x_s$ is already in the mildly nonlinear regime,
and the generating function $\varphi_s$ may already somewhat depart from its quasilinear value.
In particular, the overestimate of the negative deviation at $F \simeq 0.9$ suggests
that the negative deviation at $\delta_s \simeq -0.8$ in Fig.~\ref{fig_dPrho-z3}
was too large.

As the K-mouflage model is closer to the LCDM cosmology, in the sense that
the linear growth rate does not depend on wave number and outside of galaxies
the modification of gravity only corresponds to a small time dependence of Newton's constant,
we can expect our modeling of ${\cal P}(F)$ to fare better.
In agreement with Fig.~\ref{fig_dPrho-z3}, the modification of $\varphi_s$ has no impact
on the PDF. Thus, our prediction should be more robust than for the $f(R)$ theories.

Another factor that can explain the discrepancies found in the left panel in
Fig.~\ref{fig_dPF-z3} is our neglect of redshift-space effects.
However, since the deviations are typically smaller than the $1\sigma$ errorbars of the
observations, we do not try to extend our modeling of ${\cal P}(F)$ to redshift-space.
Indeed, the small amplitude of $\delta{\cal P}/{\cal P}$ means that the Lyman-$\alpha$
flux PDF is not a competitive probe of these modified-gravity models.

We give in Table~\ref{tab:F2} the variance $\langle \delta_F^2\rangle$ of the Lyman-$\alpha$ flux,
with $\delta_F = F/\langle F \rangle -1$.
For the modified-gravity models we use the modified generating function $\delta\varphi_s$,
which tends to overestimate the departure from the LCDM prediction.
In agreement with the previous results, the relative deviation is smallest for the K-mouflage model,
where it is only one percent.

\section{Lyman-$\alpha$ flux decrement power spectrum}
\label{sec:Lyman-flux-Pk}

We now turn to the second Lyman-$\alpha$ statistics that we consider in this paper,
the power spectrum $P_{\delta_F}({\bf k})$ of the Lyman-$\alpha$ forest flux decrement
$\delta_F = F/\langle F \rangle -1$.
Fitting formulas for $P_{\delta_F}({\bf k})$ are usually written in terms
of the power spectrum $P_{\rm L}(k)$ of the linear matter density contrast at the same
redshift, multiplied by several cutoffs and amplification factors
\cite{Arinyo-i-Prats:2015aa,Garny_2018}.
These factors account for several effects, such as the bias
between the neutral hydrogen gas distribution and the total matter distribution,
thermal broadening, redshift-space distortions, the nonlinear growth of density
fluctuations,..., and are obtained from fits to numerical simulations.
In this paper, we also use such cutoffs, which we do not accurately predict
but have realistic orders of magnitude and are fitted to simulations and observations.
However, we do not introduce an ad-hoc amplification factor and we model
the effects associated with the nonlinearity of the underlying density field by
an analytical model based on a truncated Zeldovich approximation.
This scheme cannot reach the accuracy of dedicated hydrodynamical numerical simulations,
but we can hope that it captures some of the dependence on the primordial matter
power spectrum and the growth of large-scale density perturbations.

\subsection{IGM power spectrum $P_{\rm IGM}(k)$}
\label{sec:Modeling-P-IGM}

As in section~\ref{sec:PDF} and Eq.(\ref{eq:F-delta}) for the flux PDF,
we follow the common description of the Lyman-$\alpha$ forest as due to
fluctuations in a continuous intergalactic medium (IGM)
\cite{Bi_1997,Hui:1997ac,Peirani_2014} instead of a set of discrete objects.
Thus, we first express the real-space IGM density power
spectrum of the neutral hydrogen gas in terms of the primordial matter density
power spectrum as
\be
P_{\rm IGM}(k) = P_{\rm Ztrunc}(k) \; e^{-(k/k_s)^2} ,
\label{eq:Pk-IGM}
\ee
where $P_{\rm Ztrunc}(k)$ is a truncated Zeldovich power spectrum
\cite{ZelDovich1970,Schneider1995,Taylor1996}
and $k_s$ is the smoothing wave number introduced in Eq.(\ref{eq:ks}).
We define this truncated Zeldovich power spectrum as the standard Zeldovich
power spectrum $P_{\rm Z}(k)$ associated with a truncated linear power spectrum
$P_{\rm Ltrunc}(k)$, instead of the genuine primordial linear power spectrum
$P_{\rm L}(k)$,
\be
P_{\rm Ztrunc} = \max_{k_{\rm trunc}} P_{\rm Z}[P_{\rm Ltrunc}] \;\;\; \mbox{with}
\;\;\; P_{\rm Ltrunc}(k) = P_{\rm L}(k)/(1+k^2/k_{\rm trunc}^2)^2 .
\label{eq:PL-trunc}
\ee
An alternative approach would be  to use a lognormal model for the IGM density field,
written as $\delta_{\rm IGM} \propto e^{\delta_L}$, and to use simulations to obtain
the statistical properties of this lognormal field \cite{Bi_1997}.
The advantage of our approach (\ref{eq:Pk-IGM}) is that it directly provides the
power spectrum, without the need of numerical simulations.

\begin{figure}
\begin{center}
\epsfxsize=7.5 cm \epsfysize=5.8 cm {\epsfbox{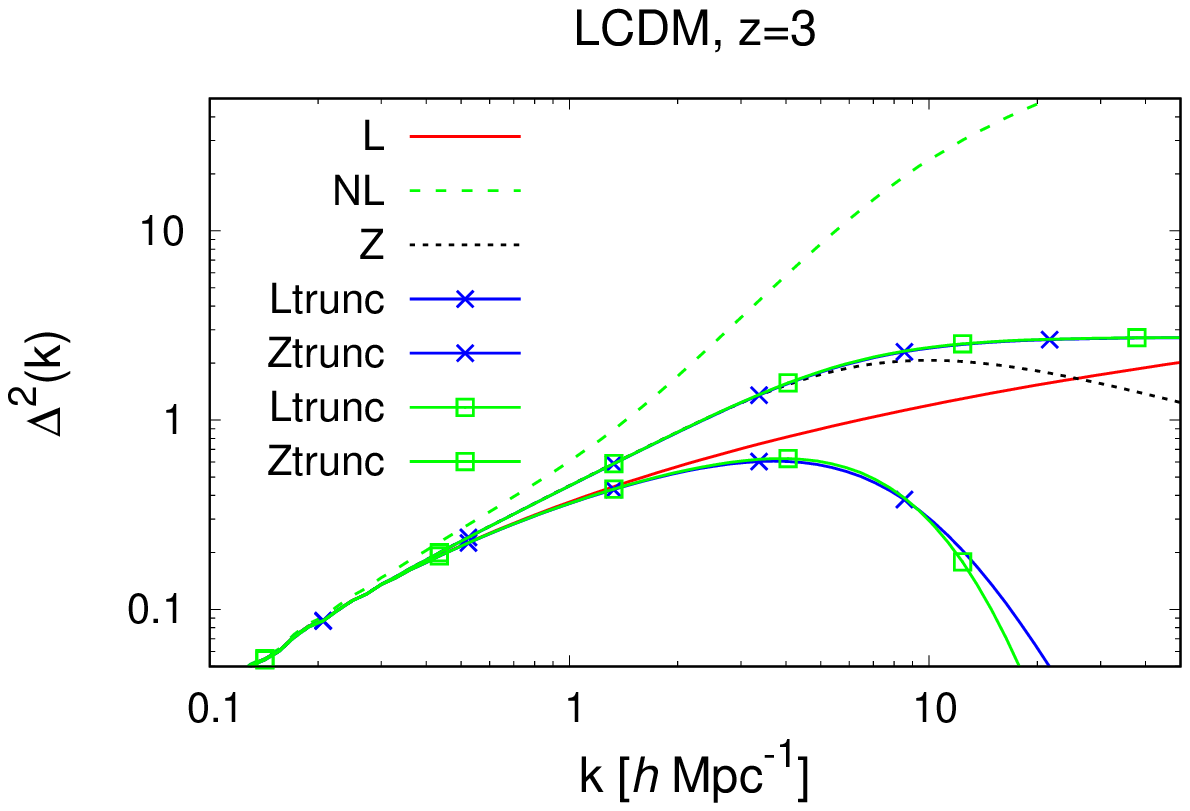}}
\epsfxsize=7.5 cm \epsfysize=5.8 cm {\epsfbox{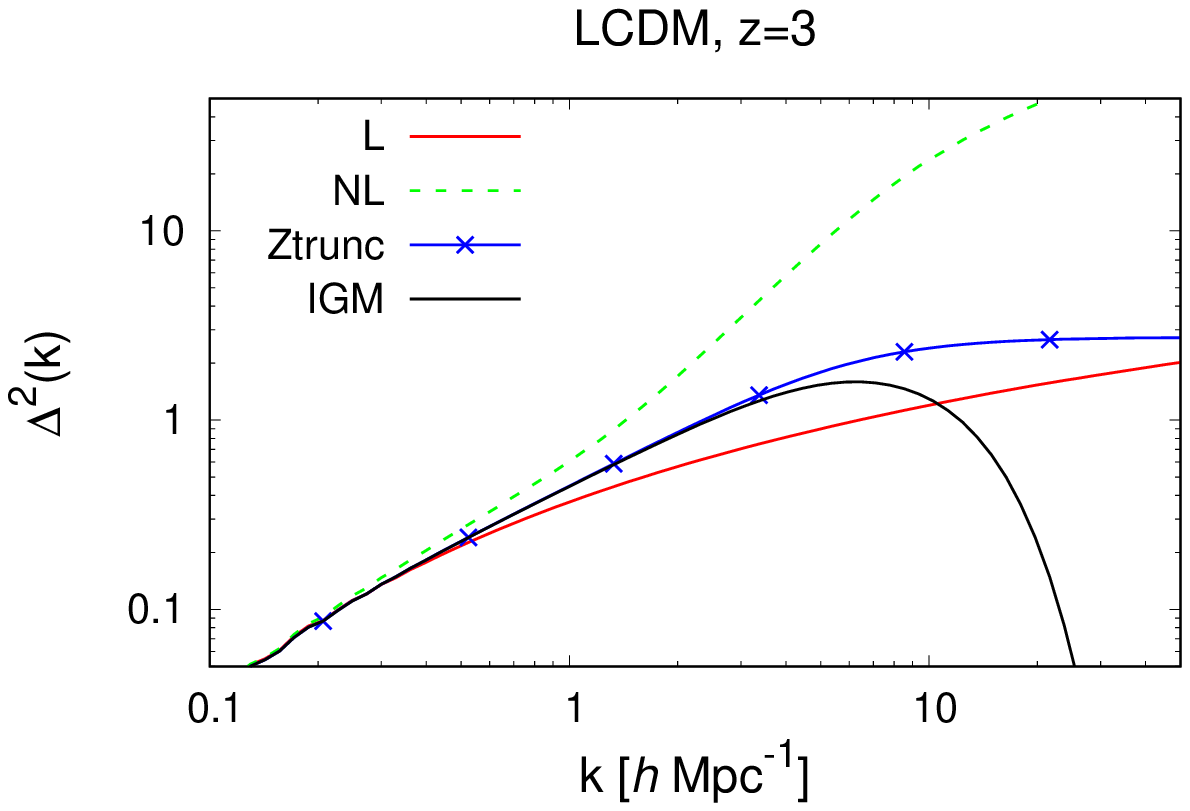}}
\end{center}
\caption{
{\it Left panel:} logarithmic power $\Delta^2(k)$ for the linear power spectrum (L),
a nonlinear model (NL), the standard Zeldovich approximation (Z),
and two truncated linear power spectra (Ltrunc) and their associated
Zeldovich approximations (Ztrunc).
{\it Right panel:} the IGM model (\ref{eq:Pk-IGM}).}
\label{fig_Pk-IGM-z3}
\end{figure}

It has been noticed for a long time that using a truncated linear power spectrum
instead of the full linear power spectrum in the Zeldovich mapping provides
a better description of large-scale structures; it actually fares better than both
the linear and lognormal approximations \cite{Coles1993}.
Indeed, the initial power at high wavenumbers gives rise to artificially large
displacements in the Zeldovich mapping, where particles simply follow their
linear trajectories. This leads to particles moving beyond collapsed structures,
instead of turning back and oscillating in gravitational potential wells,
which gives rise to a steep free-streaming cutoff of the predicted nonlinear
power spectrum, instead of the actual amplification associated with the collapse
into virialized halos.
Then, truncating the initial power at high $k$ reduces this effect and enables
one to recover the structure of the cosmic web \cite{Coles1993}.
Of course, such a scheme cannot describe the inner parts of the virialized halos.
However, this is well suited to our purposes. Indeed, Lyman-$\alpha$
forest clouds consist of mildly nonlinear density fluctuations, typically
associated with filaments or the outer parts of collapsed structures.
Therefore, removing high-density collapsed regions is actually required to focus
on the Lyman-$\alpha$ forest.
The maximization in Eq.(\ref{eq:PL-trunc}) means that the truncation
wave number $k_{\rm trunc}$ that determines $P_{\rm Ltrunc}$ is defined as
the one that maximizes $k^3 P_{\rm Ztrunc}(k)$ at high $k$.
Indeed, for large $k_{\rm trunc}$, i.e. $k_{\rm trunc} \gg k_{\rm NL}$
where $k_{\rm NL}$ is the nonlinear transition scale with
$\Delta^2_{\rm L}(k) \sim 1$
($\Delta^2 = 4 \pi k^3 P$ is the logarithmic power that also measures the
variance of density fluctuations at scale $1/k$),
we recover the primordial linear power spectrum
and the artificial smoothing of nonlinear structures.
For small $k_{\rm trunc}$, i.e. $k_{\rm trunc} \ll k_{\rm NL}$,
we already remove power in the linear regime and prevent the formation of
mildly nonlinear structures.
For $k_{\rm trunc} \sim k_{\rm NL}$, we maximize the resulting Zeldovich
power spectrum $P_{\rm Ztrunc}$, which shows a universal tail
$P_{\rm Ztrunc}(k) \propto k^{-3}$, i.e. a flat $\Delta^2_{\rm Ztrunc}(k)$
at high $k$.
This captures the self-induced truncation of the mildly nonlinear density
power spectrum we consider; the truncation is associated with the removal of
high-density virialized regions, the formation of which is set by the onset
of the nonlinear regime. This natural prescription also avoids introducing
an additional free parameter $ k_{\rm trunc}$. This also ensures that the resulting
power spectrum $P_{\rm Ztrunc}$ is not very sensitive to the form of the
cutoff $1/(1+k^2/k_{\rm trunc}^2)^\nu$, where we could as well take $\nu=1$ or 4.
Thus, we show in the left panel in Fig.~\ref{fig_Pk-IGM-z3} the power spectra
obtained without truncation and with truncation, either with $\nu=2$ (crosses)
or $\nu=4$ (squares).
We can see that at redshift $z=3$ the truncated Zeldovich approximation captures some of the
nonlinear amplification of the matter density perturbations but saturates
beyond $k_{\rm trunc} \simeq 10 h{\rm Mpc}^{-1}$, as it does not describe the inner parts
of collapsed halos. It mainly follows the standard Zeldovich approximation
up to its peak and remains constant at higher $k$. We can check that the result is
not sensitive to the exponent $\nu$ of the cutoff used for the truncation of
the linear power spectrum.

Second, the cutoff $e^{-(k/k_s)^2}$ corresponds to the damping of density
fluctuations in the gas by its nonzero pressure.
We can see in the right panel in Fig.~\ref{fig_Pk-IGM-z3} the strong
falloff at high-$k$ beyond the Jeans wave number $k_{\rm s} \sim 15 h{\rm Mpc}^{-1}$.
However, this is a relatively small-scale effect and it does not impact the linear
and weakly nonlinear growths of the IGM power spectrum.
The power spectrum $P_{\rm IGM}(k)$ shown in the right panel in Fig.~\ref{fig_Pk-IGM-z3}
corresponds to the density field $\delta_s$ with the PDF ${\cal P}(\delta_s)$
displayed in Fig.~\ref{fig_Prho-z3}, which led to the flux PDF shown in
Fig.~\ref{fig_Prho-z3}.

\subsection{Matter power spectrum for modified-gravity theories}
\label{sec:Modeling-P-IGM-modified-gravity}

\begin{figure}
\begin{center}
\epsfxsize=7.5 cm \epsfysize=5.8 cm {\epsfbox{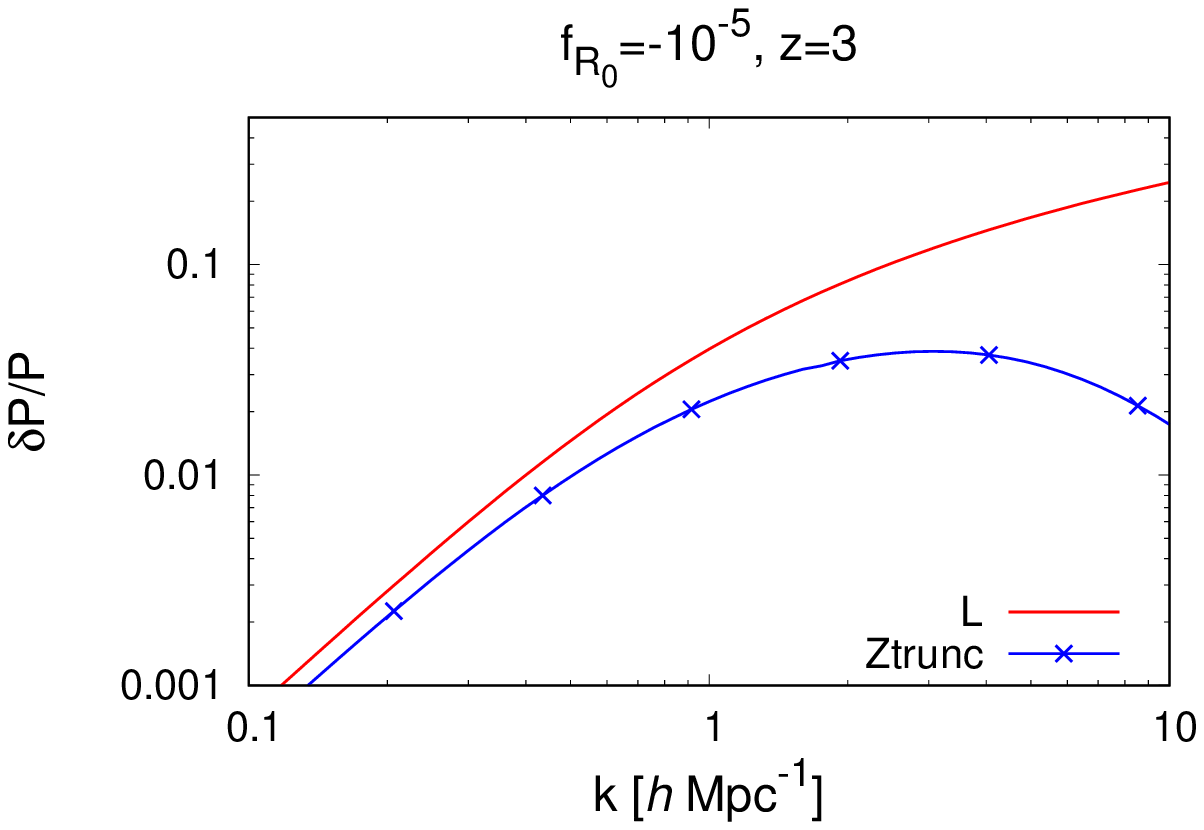}}
\epsfxsize=7.5 cm \epsfysize=5.8 cm {\epsfbox{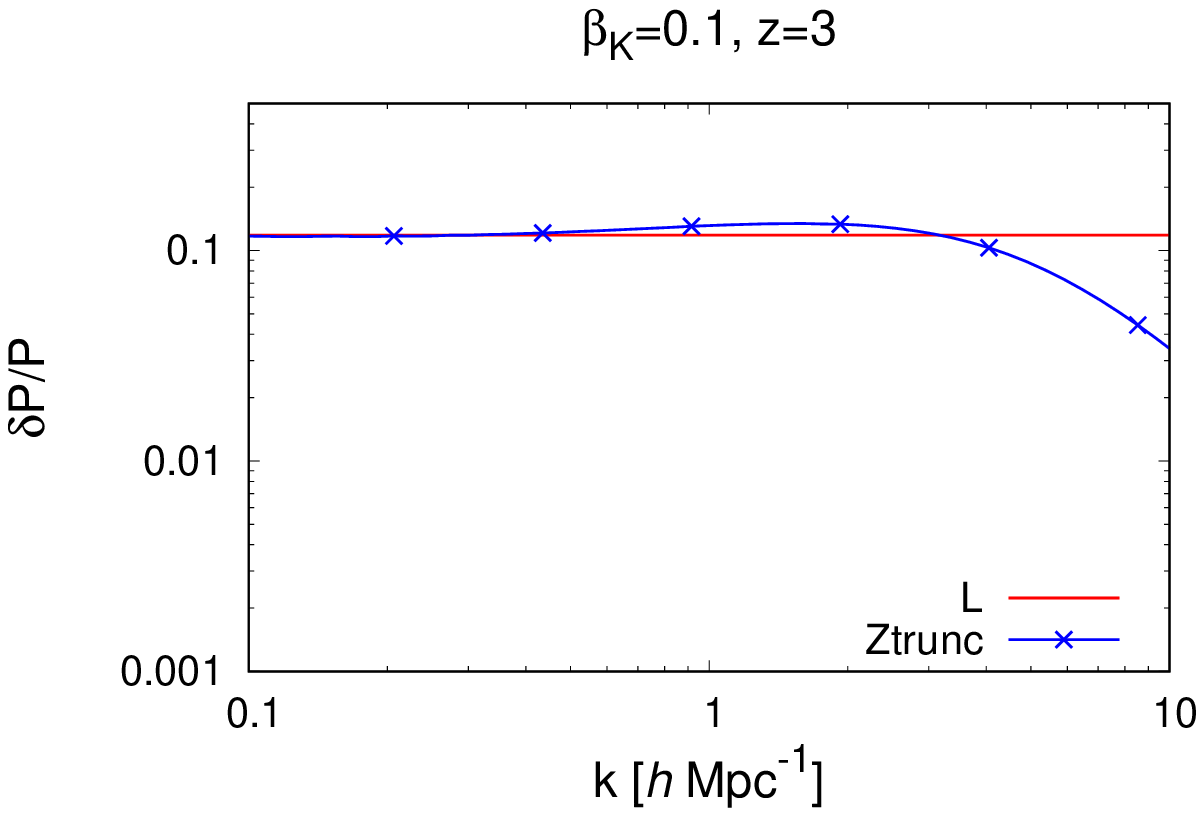}}
\end{center}
\caption{
{\it Left panel:} relative deviation from the LCDM prediction of the matter power
spectrum given by an $f(R)$ theory with $f_{R_0}=-10^{-5}$, at redshift $z=3$.
We show the linear power spectrum (L) and the truncated Zeldovich power spectrum
(Ztrunc).
{\it Right panel:} case of the K-mouflage model.
}
\label{fig_dPk-z3}
\end{figure}

\begin{figure}
\begin{center}
\epsfxsize=7.5 cm \epsfysize=5.8 cm {\epsfbox{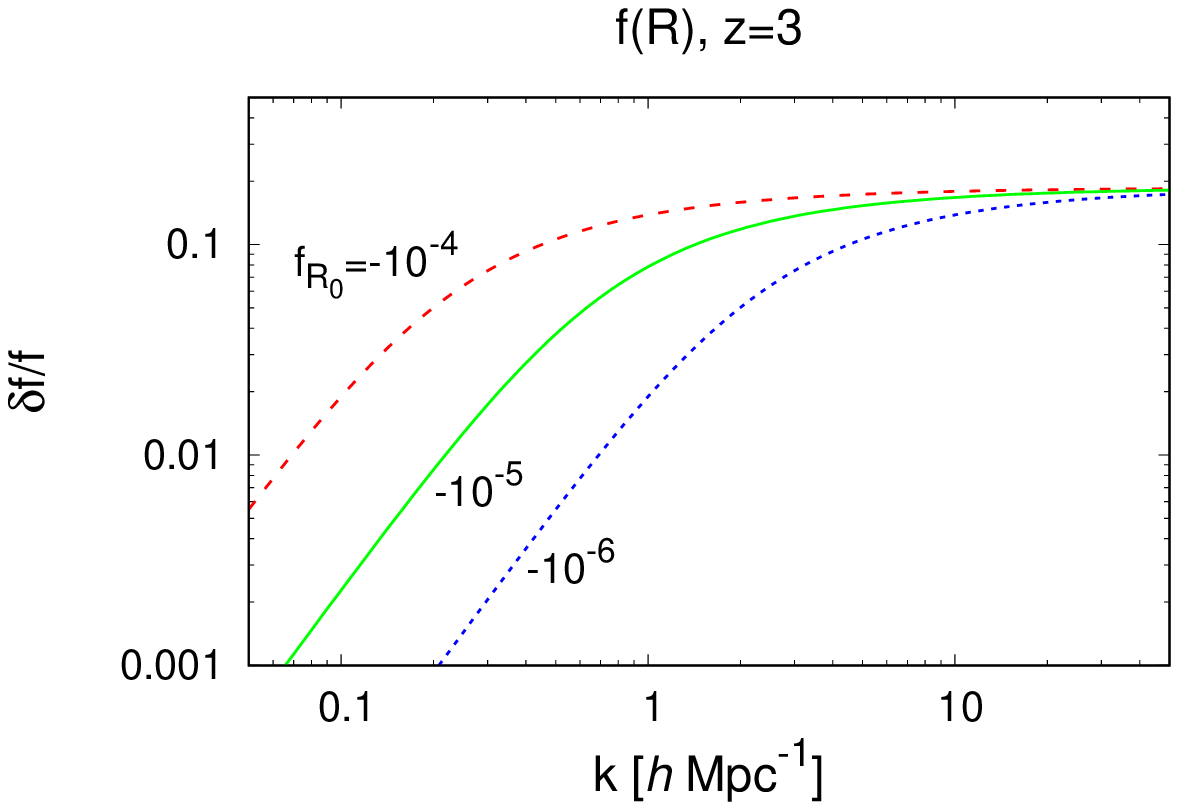}}
\epsfxsize=7.5 cm \epsfysize=5.8 cm {\epsfbox{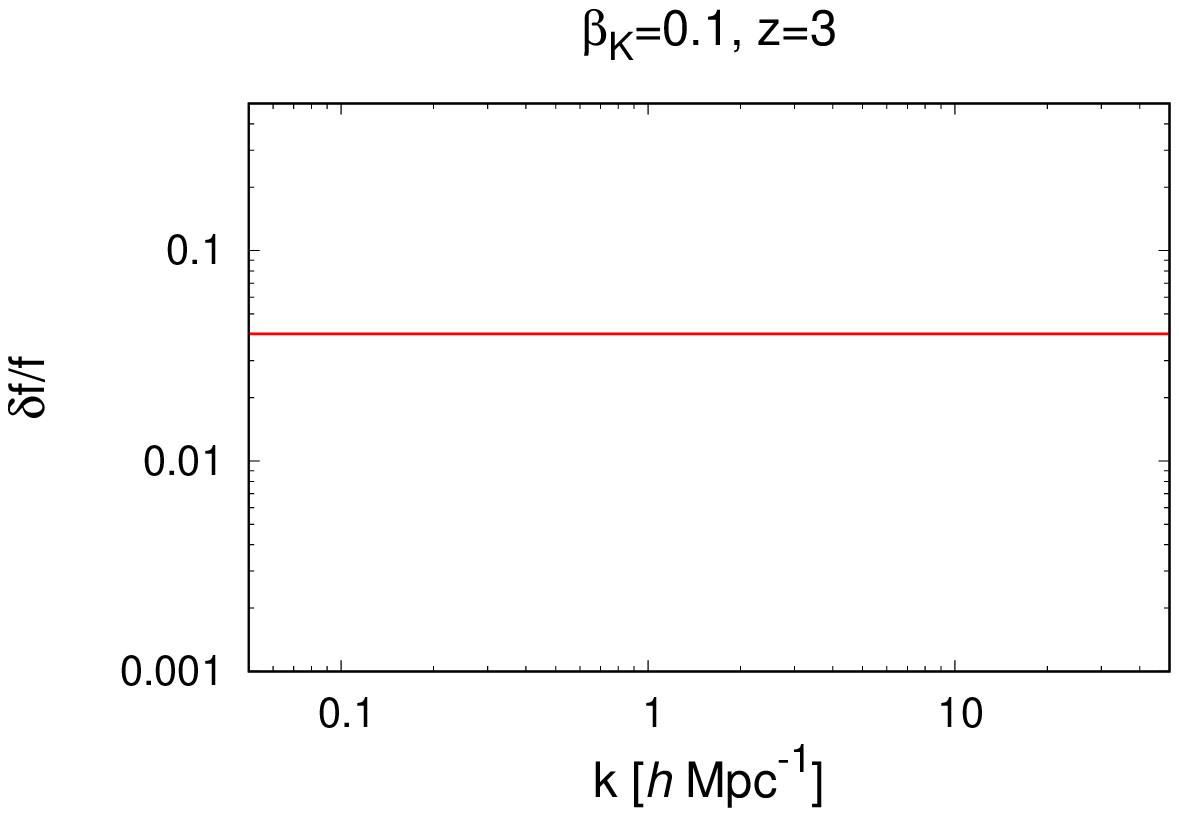}}
\end{center}
\caption{
{\it Left panel:} relative deviation from the LCDM prediction of the growth rate $f$
given by $f(R)$ theories, at redshift $z=3$.
{\it Right panel:} case of the K-mouflage model.
}
\label{fig_df-z3}
\end{figure}

We show in Fig.~\ref{fig_dPk-z3} the relative deviations of the
linear and truncated Zeldovich power spectra from the LCDM prediction.
The amplification of the growth of structure is due to the fifth force mediated
by the scalar field, and in the K-mouflage model also to the running of Newton's constant
with redshift, which now depends on the background value of the scalar field.

For the $f(R)$ theories, the relative deviation of $P_L(k)$ grows at higher $k$,
because of the mass of the scalar field which yields a characteristic scale dependence
of the linear growing mode.
Moreover, at linear order there is no chameleon screening
mechanism, which would reduce the deviation of the nonlinear power spectrum on small scales.
The deviation of the truncated Zeldovich power spectrum peaks
at the nonlinear scale and decreases at higher $k$. This is due to the
universal flat plateau already shown in Fig.~\ref{fig_Pk-IGM-z3}.
In practice, this also means that we do not need to  include explicitly the
nonlinear chameleon mechanism, as the deviation associated with nonlinear scales
is already damped.
Because we take the same IGM temperature for all cosmological scenarios,
the relative deviation of the IGM power spectrum (\ref{eq:Pk-IGM}) coincides with
that of the truncated Zeldovich power spectrum.

In contrast, for the K-mouflage scenario where the vanishing scalar-field mass
prevents any scale dependence at linear order, the relative deviation of the linear matter
power spectrum is independent of wave number.
Then, the relative deviation of the truncated Zeldovich power spectrum is constant at low $k$
and decreases again at high $k$ because of its universal plateau.

We also show in Fig.~\ref{fig_df-z3} the relative deviations of the linear growth rate
$f(k,a)$
\be
f(k,a) = \frac{\partial\ln D_+}{\partial\ln a}(k,a) .
\label{eq:f-def}
\ee
As for the linear power spectrum, the relative deviation of the growth rate $f(k,a)$
grows with $k$ for the $f(R)$ theories, while it is scale independent
for the K-mouflage model.
As explained in section~\ref{sec:f(R)}, smaller $|f_{R_0}|$ corresponds to a larger
scalar field mass $m$ and the deviation from the LCDM result is pushed to higher wave number,
while the asymptotic value (in the linear regime) is fixed and set by $\beta=1/\sqrt{6}$.
For the K-mouflage model, the constant magnitude of $\delta P/P$ and $\delta f/f$ is directly set
by the coupling constant $\beta_K$.

\subsection{Lyman-$\alpha$ power spectrum $P_{\delta_F}(k)$}
\label{sec:Modeling-P-delta-F}

We assume that the Lyman-$\alpha$ flux-decrement power spectrum
$P_{\delta_F}({\bf k},z)$ can be written in terms of the IGM density
power spectrum $P_{\rm IGM}$ as
\be
P_{\delta_F}({\bf k},z) = b_{\rm \delta_F}^2 (1+\beta \mu^2)^2 \, P_{\rm IGM}(k)
/(1+f |k\mu|/k_{\rm NL}) \, e^{-(k\mu/k_{\rm th})^2} ,
\label{eq:P-Lyman-3D}
\ee
where $\mu={\bf k}\cdot {\bf e}_z/k$ is the cosine of the wave number direction with
respect to the line of sight,
$b_{\rm \delta_F}$ the bias, $\beta$ the large-scale
anisotropy parameter associated with redshift-space distortions,
and $k_{\rm th}$ the thermal broadening cutoff wave number.

As recalled in section~\ref{sec:PDF}, even in the simplest picture the flux $F$
is a nonlinear function of the smoothed matter density, as in Eq.(\ref{eq:F-delta}).
This makes the mapping from the matter density power spectrum to the flux power spectrum
a complicated functional, and in general $P_{\delta_F}(k)$ depends on $P_L(k')$ for
a range of wave numbers $k'$. On the other hand, on large scales, nonlinear local transformations
preserve the correlation function, up to bias factors, which leads to a biased power spectrum
up to an additive constant \cite{Scherrer_1998}. This applies to the Lyman-$\alpha$
power spectrum measured on length scales that are much greater than the smoothing
scale $x_s$.
This agrees with the fact that simulations show that the complicated transformation to the
Lyman-$\alpha$ forest flux preserves the qualitative features of  dark matter
\cite{McDonald:2003aa}, which is why fitting formulae are usually written in terms of
the linear matter power spectrum, with several fitted prefactors. This is also the motivation
of Eq.(\ref{eq:P-Lyman-3D}), where $P_{\rm IGM}(k)$ converges to $P_L(k)$ on large scales
but becomes a nonlinear functional on small scales.
Then, as in standard studies \cite{McDonald:2003aa,Arinyo-i-Prats:2015aa},
we calibrate the prefactors introduced in Eq.(\ref{eq:P-Lyman-3D}) by verifying that it agrees
with numerical simulations for the 3D and 1D Lyman-$\alpha$ power spectrum of the LCDM
cosmology, and for the 1D spectrum of the $f(R)$ theories.
Then, we assume that this model also applies to the K-mouflage scenario.
This is a reasonable assumption, as we found in section~\ref{sec:PDF}
that the small-scale Lyman-$\alpha$ physics (as measured by the one-point PDF ${\cal P}(F)$)
of the K-mouflage model is even closer to the LCDM case than the $f(R)$ theories.
This is in particular seen in Tables~\ref{tab:nu2} and \ref{tab:F2}.
Moreover, all models that we consider remain very close to the LCDM cosmology.

The anisotropic $\mu$-dependent terms arise from redshift-space distortions,
due to the amplification or damping of fluctuations measured
along the line of sight because of the radial velocity fluctuations.
Indeed, the mapping from real space $\bf x$ to redshift space $\bf s$
writes as
\be
{\bf s} = {\bf x} + \frac{v_\parallel}{aH} {\bf e}_z ,
\ee
where $v_\parallel$ is the radial peculiar velocity.
Then, the velocity dispersion at a given position ${\bf x}$ redistributes the
matter at $\bf x$ over a nonzero width along the radial redshift-space
coordinate $s_{\parallel}$. This leads to a smoothing of real-space density
fluctuations and a damping of the redshift-space power spectrum at high $k$.
The factor $e^{-(k\mu/k_{\rm th})^2}$ describes the smoothing
by the termal velocity dispersion, which we take to be Gaussian
with the comoving wave number cutoff
\be
k_{\rm th} = \frac{aH}{b_{\rm th}} , \;\;\;
b_{\rm th} = \sqrt{ \frac{k_{\rm B} T}{2m_p} } ,
\ee
where $b_{\rm th}$ is the thermal velocity dispersion \cite{Garzilli:2015aa}.
The factor $1/(1+f | k\mu | /k_{\rm NL})$ describes the smoothing by the velocity
dispersion due to the virialization of collapsed structures. On nonlinear scales,
beyond $k_{\rm NL}$, shell crossing appears and different velocity streams coexist
at the same physical space location $\bf x$. This must again damp the redshift-space
power spectrum.
The factor $f$ expresses that this damping appears earlier when the linear velocity
perturbations are amplified with respect to the linear density field.
The factor $(1+\beta \mu^2)^2$ is the usual Kaiser effect \cite{Kaiser_1987},
which describes that on large linear scales the single-stream velocity field
amplifies the density perturbations, as matter is moving inward onto overdense regions.
We simply take $\beta \simeq1.3 \; f$, where $f(k,a)$ is the linear growth rate
defined in Eq.(\ref{eq:f-def}).
In principles, the factor $\beta$ is defined as $\beta = f b_{\delta_F,\eta} / b_{\delta_F,\delta}$,
where we distinguish the biases with respect to the linear density and velocity fields,
$b_{\delta_F,\delta}=\partial \delta_F / \partial\delta$ and
$b_{\delta_F,\eta}=\partial \delta_F / \partial\eta$, with
$\eta = - (\partial v_\parallel/\partial x_{\parallel})/(aH)$
\cite{Seljak_2012,Arinyo-i-Prats:2015aa}.
At lowest order, these biases may be computed from mode couplings such as
Eq.(\ref{eq:nu2-def}).
However, we found that the analytical models for $b_{\delta_F,\delta}$
and $b_{\delta_F,\eta}$ \cite{Seljak_2012,Cieplak_2016}
do not fare very well. They do not improve the agreement
with numerical simulations and are not very stable, in particular the large
inaccuracies on $b_{\delta_F,\eta}$ can lead to artificially large or small
values for $\beta$.
This agrees with the results of \cite{Cieplak_2016}, who pointed out that velocity effects
and redshift-space distortions are very difficult to capture by simple analytical models.
Therefore, we keep the simple expression $\beta \sim 1.3 f$, which appears to
be more robust and agrees with numerical simulations at redshift $z\simeq 3$
\cite{Arinyo-i-Prats:2015aa}.
The very small departure of $\nu_2$ from the LCDM result found in Table~\ref {tab:nu2},
especially for the K-mouflage model, suggests that this should be a reasonable approximation
for the modified-gravity models we study in this paper.
The prefactor $b_{\rm \delta_F}^2$ is fitted to the observations.
Apart from direct hydrodynamical simulations, an alternative would be to simulate the
density and velocity fields associated with the truncated Zeldovich approximation.
which allows a more accurate treatment of thermal and redshift-space distortions
\cite{Hui:1997ac}. However, as we only wish to estimate the magnitude of the impact
of modified-gravity theories, for simplicity we keep the analytical model (\ref{eq:P-Lyman-3D}).
For precise measurements, one should in any case develop dedicated hydrodynamical
simulations \cite{Rossi:2014aa,Peirani_2014,Lochhaas_2016}.

\begin{figure}
\begin{center}
\epsfxsize=7.5 cm \epsfysize=5.8 cm {\epsfbox{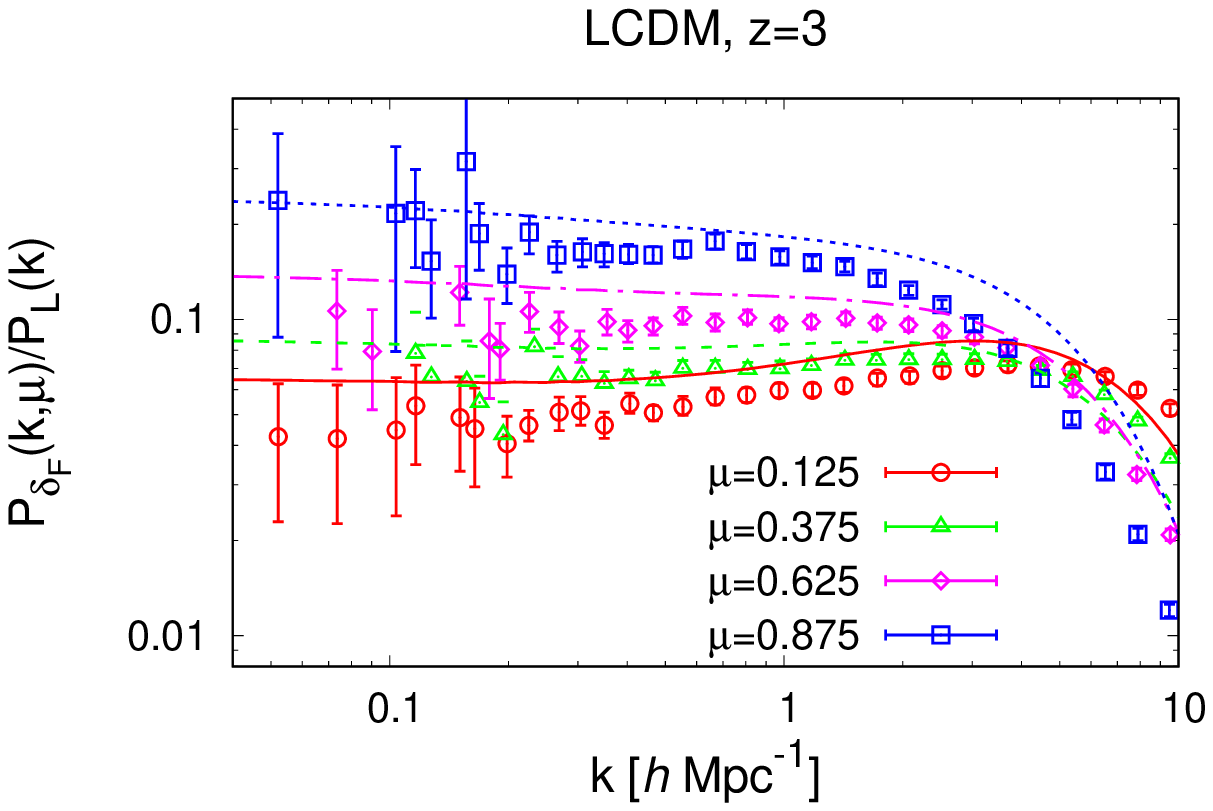}}
\epsfxsize=7.5 cm \epsfysize=5.8 cm {\epsfbox{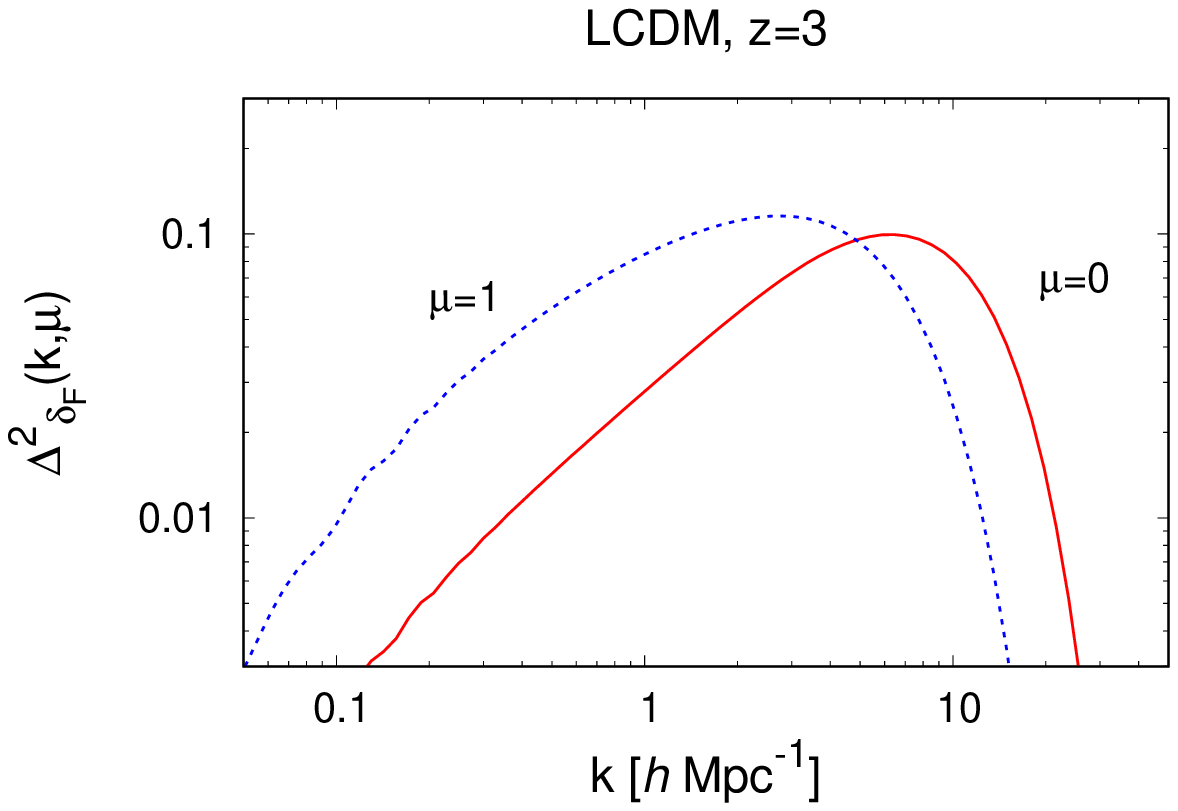}}
\end{center}
\caption{
{\it Left panel:} ratio $P_{\delta_F}(k,\mu)/P_L(k)$
for $\mu=0.125$, $0.375$, $0.625$ and $0.875$, from bottom to top.
The symbols are the results of numerical simulations \cite{Arinyo-i-Prats:2015aa}.
{\it Right panel:} logarithmic power spectrum $\Delta^2_{\delta_F}$
along the line of sight and perpendicular to the line of sight.
}
\label{fig_rPk-Lyman-z3}
\end{figure}

We show in the left panel in Fig.~\ref{fig_rPk-Lyman-z3} the ratio of
the Lyman-$\alpha$ power spectrum to the linear matter density power spectrum,
at redshift $z=3$ as a function of the wave number $k$, for several values of $\mu$.
In agreement with Eq.(\ref{eq:P-Lyman-3D}), higher values of $\mu$
(i.e. directions increasingly parallel to the line of sight) amplify the
power spectrum on large scales, because of the Kaiser effect, and damp the power
on small scales because of the $\mu$-dependent cutoffs, due to the smoothing
by the velocity dispersion that arises from the thermal distribution and
the gravitational multistreaming.
The agreement with the numerical simulations \cite{Arinyo-i-Prats:2015aa}
is not perfect, as expected for
such a simple model as (\ref{eq:P-Lyman-3D}), but we recover the main trends
and the magnitude of these redshift-space distortions.
This suggests that our model captures the main processes at work.
We show in the right panel in Fig.~\ref{fig_rPk-Lyman-z3} the logarithmic power
spectrum, $\Delta^2_{\delta_F} = 4\pi k^3 P_{\delta_F}(k,\mu)$ for $\mu=1$ and $\mu=0$.
In agreement with the left panel, the redshift-space distortions amplify the
power at low $k$ but give rise to a steeper falloff at high $k$.

\begin{figure}
\begin{center}
\epsfxsize=7.5 cm \epsfysize=5.8 cm {\epsfbox{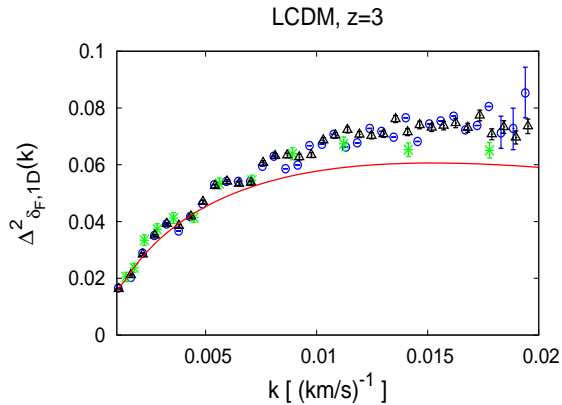}}
\end{center}
\caption{
Logarithmic 1D power spectrum $\Delta^2_{\delta_F,{\rm 1D}}$.
The data points are observations from \cite{McDonald:2006aa} (stars),
\cite{Palanque-Delabrouille:2013aa} (circles) and \cite{Chabanier2018} (triangles).
The solid line is our model.
}
\label{fig_Pk-Lyman-1D}
\end{figure}

The expression (\ref{eq:P-Lyman-3D}) gives the anisotropic 3D Lyman-$\alpha$
power spectrum, over all directions of $\bf k$.
The observed 1D power spectrum along the line of sight is given by the standard
integral
\be
P_{\delta_F,{\rm 1D}}(k_z) = \int_{-\infty}^{\infty} dk_x dk_y P_{\delta_F}({\bf k})
= 2\pi  \int_{k_z}^{\infty} dk \, k P_{\delta_F}(k,\mu=k_z/k) .
\label{eq:P-Lyman-1D}
\ee
This also defines the 1D logarithmic power as
$\Delta^2_{\delta_F,{\rm 1D}}(k)=(k/\pi) P_{\delta_F,1D}(k)$,
which we compare with observations 
\cite{McDonald:2006aa,Palanque-Delabrouille:2013aa,Chabanier2018}
in Fig.~\ref{fig_Pk-Lyman-1D}.
In agreement with Fig.~\ref{fig_rPk-Lyman-z3}, we recover the broad shape of the
observed 1D Lyman-$\alpha$ power spectrum.
The amplitude itself is not predicted, as the bias $b_{\delta_F}$ is fitted
to these observations.
The lack of power at high $k$, $k \gtrsim 0.015 \,
({\rm km/s})^{-1}$ suggests some
tension between the observations and the numerical simulations \cite{Arinyo-i-Prats:2015aa},
as increasing the power at high $k$ of the model would then worsen the agreement
with the numerical simulations shown in Fig.~\ref{fig_rPk-Lyman-z3}.
We do not tune our model to  fit better the observations, to keep a reasonable
agreement with both simulations and observations. This is likely to give a more
robust framework. A more accurate modeling would require detailed comparisons
between observations and simulations to better understand the different physical effects
that enter the transformation from the linear matter density power spectrum to the
Lyman-$\alpha$ power spectrum.

\subsection{Lyman-$\alpha$ power spectrum for modified-gravity theories}
\label{sec:P-Lyman-modified-gravity}

\begin{figure}
\begin{center}
\epsfxsize=7.5 cm \epsfysize=5.8 cm {\epsfbox{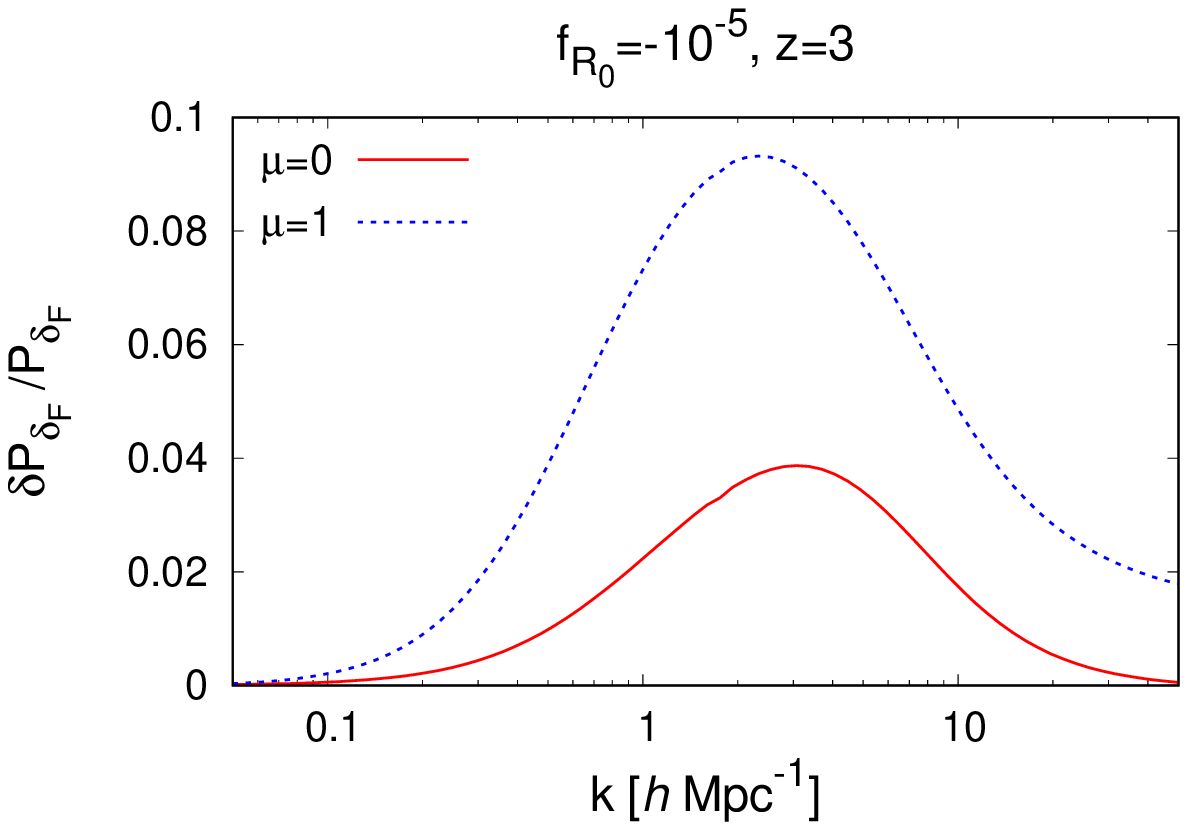}}
\epsfxsize=7.5 cm \epsfysize=5.8 cm {\epsfbox{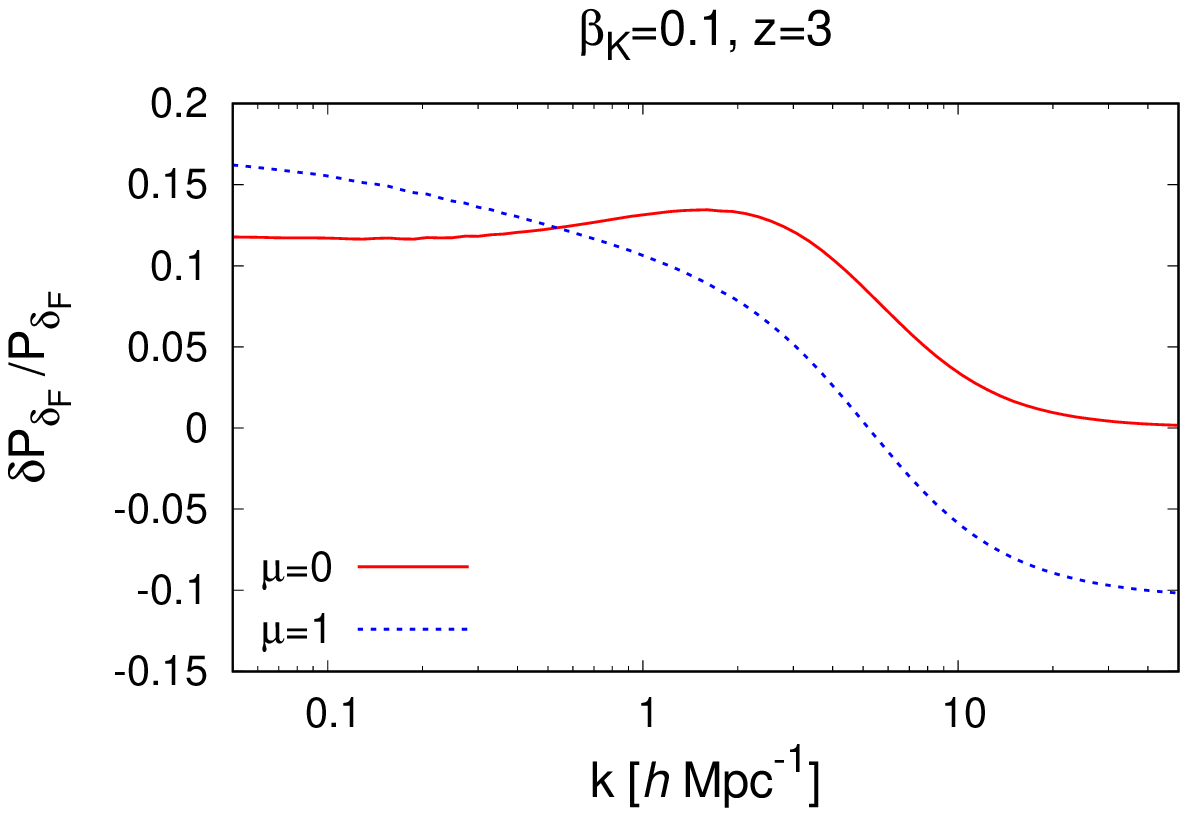}}
\end{center}
\caption{
{\it Left panel:} relative deviation from the LCDM prediction of the 3D Lyman-$\alpha$
power spectra given by an $f(R)$ theory with $f_{R_0}=-10^{-5}$, at redshift $z=3$,
along directions orthogonal and parallel to the line of sight.
{\it Right panel:} case of the K-mouflage model.
}
\label{fig_dPk-Lyman-3D-z3}
\end{figure}

We show in Fig.~\ref{fig_dPk-Lyman-3D-z3} the deviations
with respect to the LCDM prediction for the 3D Lyman-$\alpha$ power spectra.
For the $f(R)$ theories, we can see in the left panel that on large linear and weakly
nonlinear scales the relative deviation of the Lyman-$\alpha$ power spectrum grows with $k$,
following the rise of the modification to the matter power spectrum itself.
The relative deviation is greater along the radial direction, which is also sensitive
to the modification of the redshift-space factor $f$.
The relative deviation of the transverse power spectrum decreases at higher $k$,
following the behavior of the truncated Zeldovich power spectrum.
Along the radial direction, the relative deviation does not decrease at high $k$
and goes to a finite value. This is because it remains set by the change of the
overall prefactor $(1+\beta\mu^2)^2$ in Eq.(\ref{eq:P-Lyman-3D}), through the
modification of the growth rate $f$.
However, this result should not be trusted at nonlinear scales,
$k \gtrsim 1 h/{\rm Mpc}$, because this simple form of the Kaiser amplification
factor only holds on linear scales.
However, this range does not dominate the integral (\ref{eq:P-Lyman-1D}) that
gives the 1D Lyman-$\alpha$ power spectrum.

For the K-mouflage model, the relative deviation of the Lyman-$\alpha$ power spectrum
is scale independent on large linear and weakly nonlinear scales,
as it is set by the relative deviation of the linear matter power spectrum.
As for the $f(R)$ scenarios, the relative deviation of the transverse power
spectrum decreases at high $k$, following the behavior of the truncated Zeldovich
power spectrum.
Along the radial direction, the relative deviation shows a faster decrease
and even becomes negative at high $k$ because of the numerator in
Eq.(\ref{eq:P-Lyman-3D}), associated with the greater velocity dispersion.
Again, this behavior should not be trusted as these scales are already
in the highly nonlinear regime, which is not expected to be well described
by our simple modeling.

\begin{figure}
\begin{center}
\epsfxsize=7.5 cm \epsfysize=5.8 cm {\epsfbox{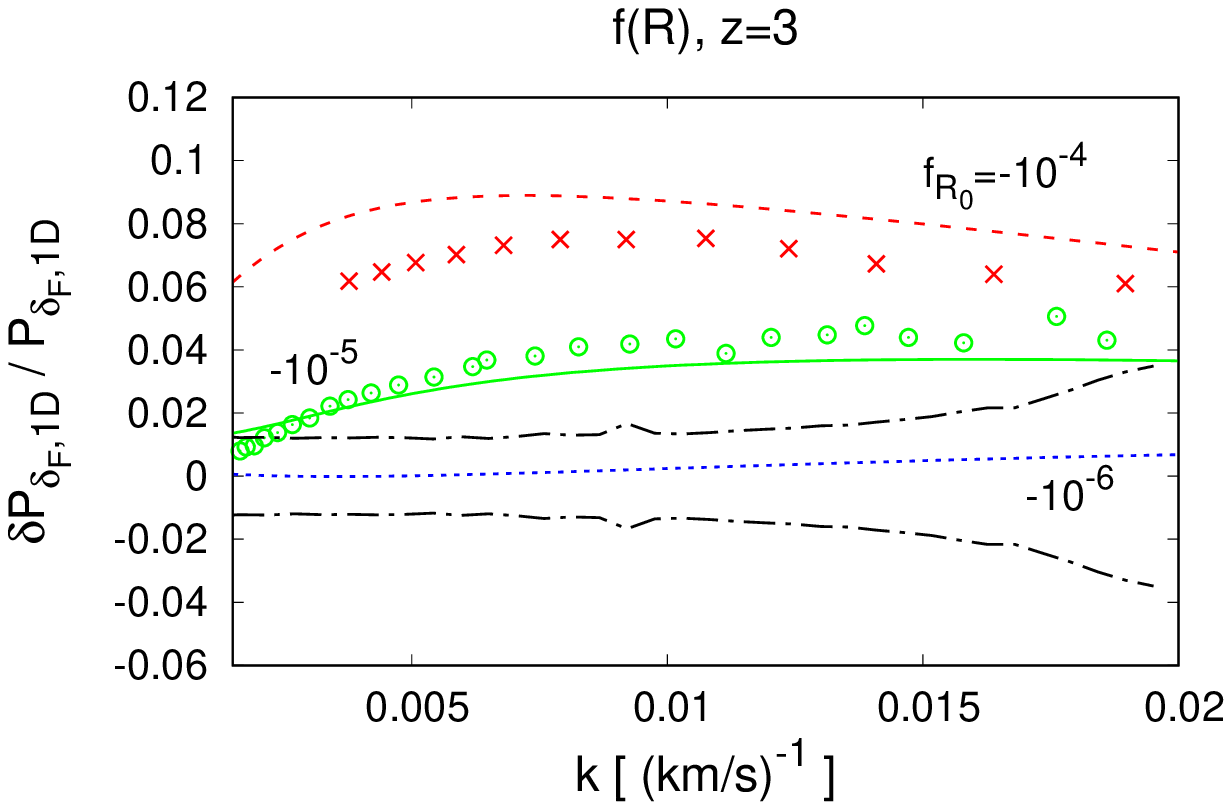}}
\epsfxsize=7.5 cm \epsfysize=5.8 cm {\epsfbox{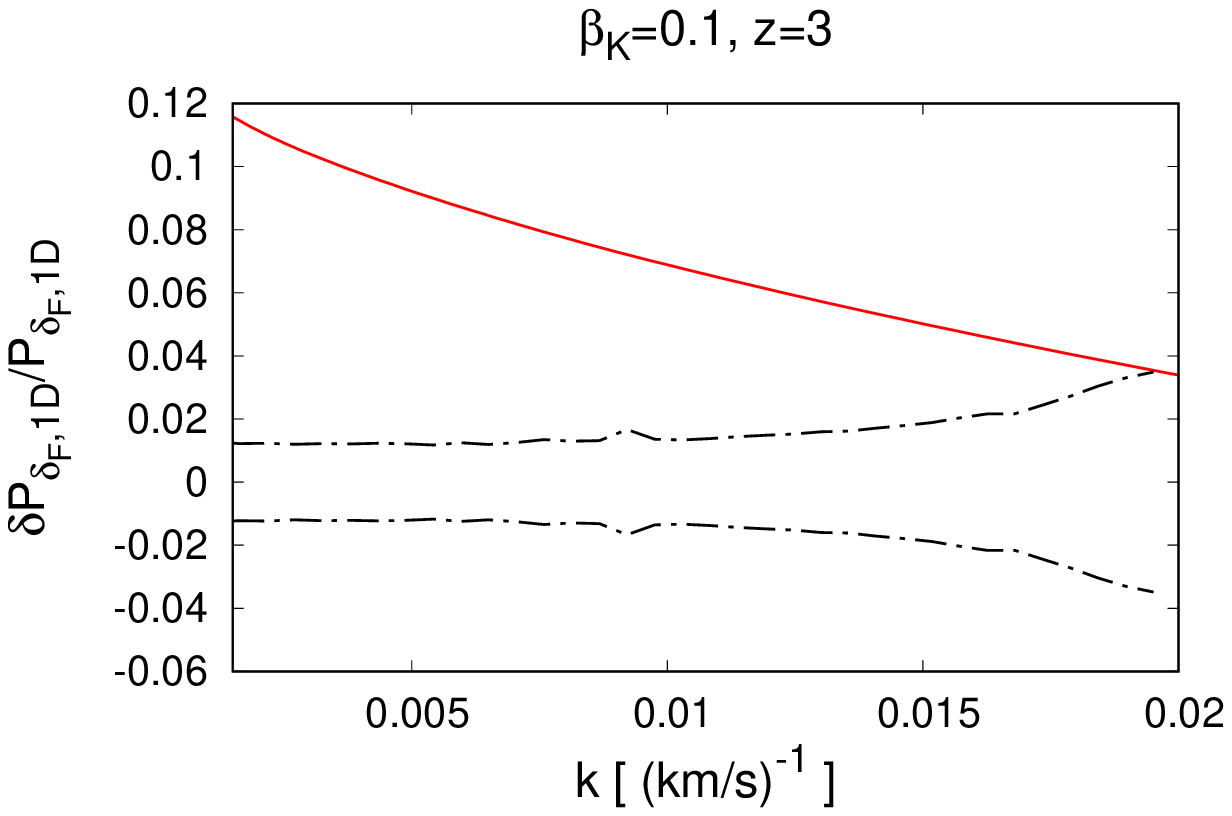}}
\end{center}
\caption{
{\it Left panel:} relative deviation from the LCDM prediction of the 1D Lyman-$\alpha$
power spectrum at $z=3$ given by $f(R)$ theories, with $f_{R_0}=-10^{-4}$ (red dashed line),
$-10^{-5}$ (green solid line) and $-10^{-6}$ (blue dotted line).
The points are the numerical simulations of \cite{Arnold2015} for
$f_{R_0}=-10^{-4}$ (red crosses) and $f_{R_0}=-10^{-5}$ (green circles).
The symmetric upper and lower black dot-dashed lines are the $\pm 1\sigma$
relative errors of the observational results of \cite{Chabanier2018}.
{\it Right panel:} case of the K-mouflage model.
}
\label{fig_dPk-Lyman-1D-z3}
\end{figure}

We show in Fig.~\ref{fig_dPk-Lyman-1D-z3} the relative deviation of the
1D Lyman-$\alpha$ power spectrum.
As compared with the 3D power spectra displayed in Fig.~\ref{fig_dPk-Lyman-3D-z3},
the integration over the transverse wave numbers smoothes the
relative deviation from the LCDM prediction. Thus, we obtain a deviation of order
$4\%$ for $f_{R_0}=-10^{-5}$, which does not vary much over
$0.005 < k < 0.02 \, ({\rm km/s})^{-1}$, and a deviation of order
$7\%$ for $f_{R_0}=-10^{-4}$.
Our results agree reasonably well with the numerical simulations
from \cite{Arnold2015}, which suggests that the model captures the main
dependence on the cosmology.
The modest value of the deviation from the LCDM cosmology and the lack of salient
features suggest that the Lyman-$\alpha$ power spectrum is not a competitive tool
to constrain these $f(R)$ theories, which are already strongly constrained
by astrophysical probes and Solar System tests of gravity that imply
$| f_{R_0} | \lesssim 10^{-6}$.
Thus, it appears that to obtain useful constraints on these scenarios one
needs to reconstruct the 3D power spectrum, shown in Fig.~\ref{fig_dPk-Lyman-3D-z3},
which shows a stronger scale dependence and a higher magnitude for the peak
of the deviation from the LCDM power spectrum.

For the K-mouflage model, the 1D Lyman-$\alpha$ flux power spectrum shows
a smooth relative deviation that slowly decreases with $k$.
This is because of the scale independence for the relative deviation of the linear matter
power spectrum, due to the zero mass of the scalar field, while at high $k$ nonlinear
effects come into play that somewhat damp the dependence of the flux power spectrum
on the underlying linear power spectrum.
The comparison with the $1\sigma$ relative error of the observational results of
\cite{Chabanier2018} indicates that a precise analysis could constrain
K-mouflage models at the level of $\beta_K \lesssim 0.1$.
This can be compared for instance with CMB and background constraints, which give
$\beta_K \lesssim 0.2$ \cite{Benevento2018}.
Therefore, in contrast with the case of the $f(R)$ theories,
the Lyman-$\alpha$ power spectrum could provide competitive constraints
for these models.
This is partly due to their different screening mechanisms.
In the case of K-mouflage models, the nonlinear screening that ensures convergence
to General Relativity in the Solar System has not impact on weakly nonlinear cosmological scales
(because this corresponds to different regimes of the kinetic function $K(\chi)$
that are not necessarily related), and the tests of gravity in the Solar System or astrophysical
environments only imply $\beta_K \lesssim 0.1$ (provided $K'(\chi)$ is sufficiently large
in the small-scale quasistatic regime).

Obtaining competitive constraints would require a more accurate modeling, or at least
a comparison with a set of K-mouflage numerical simulations to check the accuracy
of our modeling, which we leave to future works.
On the other hand, the rather large deviation found in Fig.~\ref{fig_dPk-Lyman-1D-z3},
as compared with the small deviation of the one-point variance $\langle \delta_F^2 \rangle$
found in Table~\ref{tab:F2}, suggests that our result is robust and would not be
removed by the impact on the small-scale IGM physics.

The comparison with the case of the $f(R)$ theories also shows that the shape
of the relative deviation of the Lyman-$\alpha$ flux power spectrum can provide useful
constraints on the mass of the scalar field, or more generally on whether new length scales
are introduced by a possible nonstandard cosmological scenario.

\subsection{Degeneracies with physical parameters}
\label{sec:parameters}

\begin{figure}
\begin{center}
\epsfxsize=7.5 cm \epsfysize=5.8 cm {\epsfbox{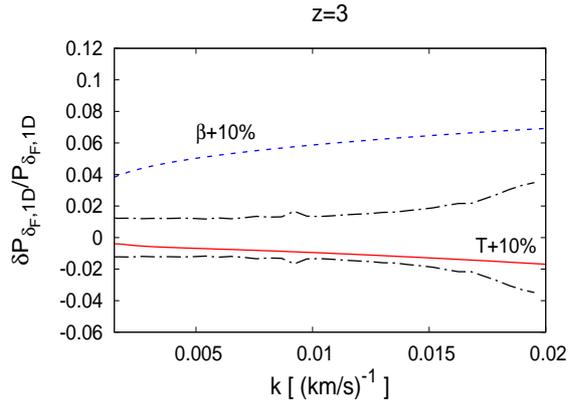}}
\end{center}
\caption{
Relative deviation of the 1D Lyman-$\alpha$ power spectrum at $z=3$ due to
a $10\%$ increase of the IGM temperature (red solid line) or a $10\%$ increase of the
redshift-space bias factor $\beta$ (blue dashed line).
The symmetric upper and lower black dot-dashed lines are the $\pm 1\sigma$
relative errors of the observational results of \cite{Chabanier2018}.
}
\label{fig_param_dPk-Lyman-1D-z3}
\end{figure}

To investigate the robustness of our results with respect to small-scale modifications
of the IGM physics, we show in Fig.~\ref{fig_param_dPk-Lyman-1D-z3} the relative
deviations of the 1D Lyman-$\alpha$ power spectrum for a $10\%$ increase of
the IGM temperature $T$ or of the bias ratio $\beta$.
The increased temperature implies a stronger damping of the IGM power spectrum
(\ref{eq:Pk-IGM}), because of the greater Jeans length scale, and of the Lyman-$\alpha$
power spectrum (\ref{eq:P-Lyman-3D}), because of the greater thermal broadening.
As seen in Fig.~\ref{fig_param_dPk-Lyman-1D-z3}, this yields a flat decrease of
$P_{\delta_F,1D}(k)$ of about $1\%$.
The increased factor $\beta$ yields a flat increase of $P_{\delta_F,1D}(k)$ of about $6\%$,
through Eq.(\ref{eq:P-Lyman-3D}).
The comparison with Fig.~\ref{fig_dPk-Lyman-1D-z3} shows that the impact of these
physical parameters remains modest and gives a scale dependence that is different
from the modified-gravity models. This shows that it is possible to break degeneracies
between such effects and modified gravity by using the scale dependence
of the power spectrum.

\section{Conclusion}
\label{sec:conclusion}

In this paper, we have presented a simple modeling of the Lyman-$\alpha$ forest
statistics to estimate the impact of two modified-gravity scenarios, the $f(R)$ and K-mouflage
models.

We have first considered the probability distribution function ${\cal P}(F)$
of the Lyman-$\alpha$ flux.
We find that this is not a very competitive probe of modified-gravity theories.
It is also difficult to model analytically for scenarios that introduce a new scale dependence,
such as the $f(R)$ theories, where the deviation of the PDF of the matter density contrast
goes somewhat beyond the amplification of the variance and involves the higher-order
cumulants.

Next, we have developed a simple modeling of the Lyman-$\alpha$ flux power spectrum,
using a truncated Zeldovich approximation. This provides
a good starting point to describe the Lyman-$\alpha$ power spectrum
at $z \sim 3$, as it captures weakly nonlinear structures while removing
the contributions of highly nonlinear objects that do not correspond to
Lyman-$\alpha$ forest clouds.
Taking into account thermal and redshift-space effects, through bias and cutoff
factors, we obtain a reasonably good agreement with numerical simulations
of the concordance LCDM model and with observations.
For the $f(R)$ models, where numerical simulations are available,
we also obtain a reasonable agreement with the numerical results.
We find that because of the line of sight integration, the deviations
from the LCDM prediction for the 1D Lyman-$\alpha$ power spectrum are modest
and flat over the observed range of wave numbers.
This will make it difficult to derive competitive constraints for these models,
which are already very strongly constrained by astrophysical probes.
In contrast, because of their different screening mechanism, the K-mouflage models
are less strongly constrained by astrophysical probes, which only constrain
the negative-$\chi$ range of $K(\chi)$ with the bound $\beta_K \lesssim 0.1$.
There, we find that the Lyman-$\alpha$ power spectrum could provide competitive
constraints as compared with CMB and background measurements.
However, this would require numerical simulations to check the accuracy of the
analytical modeling.

In addition, the 3D Lyman-$\alpha$ power spectrum shows a stronger scale dependence,
which is sensitive to the details of the modified-gravity theory,
in particular to its scale dependence (which typically arises from the scale
associated with the mass of the new scalar field).
Therefore, reconstructing the 3D power spectrum by correlating neighboring
lines of sight \cite{Pichon_2001,Slosar_2011,Cisewski_2014,Ozbek_2016,Font-Ribera:2018aa}
could provide a useful probe of alternative cosmologies.

\section*{Acknowledgments}
This work is supported in part by the EU Horizon 2020 research and innovation programme under the Marie-Sklodowska grant No. 690575. This article is based upon work related to the COST Action CA15117 (CANTATA) supported by COST (European Cooperation in Science and Technology).

\appendix1

\section{Cumulant generating function}
\label{app:generating-function}

We introduced in Eq.(\ref{eq:phi-def}) the generating function $\varphi(y)$ of the
cumulants of the density field.
The first expression is actually the inverse Laplace transform of $e^{-\varphi/\sigma^2}$,
which means that $\varphi(y)$ is also defined as the Laplace transform of the PDF
${\cal P}(\delta)$,
\be
e^{-\varphi(y)/\sigma^2} = \int_{-1}^{\infty} d \delta \, e^{-y\delta/\sigma^2} {\cal P}(\delta)
= \langle e^{-y\delta/\sigma^2} \rangle
= \int {\cal D}\delta_L e^{-y\delta[\delta_L]/\sigma^2} e^{-\frac{1}{2} \delta_L \cdot C_L^{-1} \cdot
\delta_L} .
\ee
In the last expression we wrote the average $\langle e^{-y\delta/\sigma^2} \rangle$
as a statistical path integral over the initial conditions, defined by the Gaussian linear density
field $\delta_L({\bf x})$ with its two-point correlation $C_L$.
As described in \cite{Valageas2002,BraxPV2012}, in the limit $\sigma\to 0$
this integral is dominated by a spherically symmetric saddle point.
This gives $\varphi(y)$ as the minimum
\be
\varphi(y) = \min_{\delta_{Lq}} \left[ y {\cal F}(\delta_{Lq}) + \frac{1}{2} \frac{\sigma_x^2}{\sigma_q^2}
\delta_{Lq}^2 \right] , \;\;\; q = (1+\delta)^{1/3} x ,
\label{eq:phi-Legendre}
\ee
see also \cite{Uhlemann_2016,Bernardeau_2016,Ivanov2018} for recent works.
Here $\delta(x) = {\cal F}(\delta_{Lq})$ is the nonlinear density contrast at the scale $x$ reached by
the collapse of a linear density contrast $\delta_{Lq}$ at the Lagrangian scale $q$,
and $q = (1+\delta)^{1/3} x$ expresses the conservation of matter.
The functional $\delta[\delta_L]$ is reduced to the spherically symmetric function
${\cal F}(\delta_{Lq})$ because for scale-independent gravity theories, such as General
Relativity and the K-mouflage model (in its unscreened regime), different shells decouple
in the spherical gravitational dynamics before shell crossing.
The $f(R)$ theories introduce a scale dependence that implies that the dynamics of all shells
are coupled. This makes the problem much more complex, but as in \cite{BraxPV2012}
we use a simple approximation where we only follow the collapse of the mass scale $q$
and take a fixed ansatz for the radial profile at each time step.
Then, the generating function $\varphi$ associated with any modified-gravity scenario is
determined by the new spherical-collapse mapping ${\cal F}(\delta_L)$,
see \cite{BraxPV2012} for details.

\section{Second-order perturbation theory}
\label{app:second-order}

Expanding the matter density contrast $\delta$ in powers of the linear density
contrast $\delta_{L0}$ extrapolated to $z=0$, we can write the second-order term
in Fourier space as
\bea
\delta^{(2)}({\bf k},\eta) & = & \int_{-\infty}^{\eta} d\eta'
\int d{\bf k}_1 d{\bf k}_2 \, \delta_D({\bf k}_1+{\bf k}_2-{\bf k})
\delta_{L0}({\bf k}_1) \delta_{L0}({\bf k}_2)
\biggl [ R_{11}(k;\eta,\eta') \alpha({\bf k}_1,{\bf k}_2) \nonumber \\
&& \times \frac{\partial D_+}{\partial\eta}(k_1,\eta') D_+(k_2,\eta')
+ R_{12}(k;\eta,\eta') \biggl ( \beta({\bf k}_1,{\bf k}_2)
\frac{\partial D_+}{\partial\eta}(k_1,\eta') \frac{\partial D_+}{\partial\eta}(k_2,\eta')
\nonumber \\
&& + \gamma^s_{2;11}({\bf k}_1,{\bf k}_2;\eta') D_+(k_1,\eta') D_+(k_2,\eta')
\biggl ) \biggl ] ,
\eea
where we use the notations of \cite{BraxPV2013}.
Here $\eta=\ln a$, $D_+$ is the linear growing mode, which depends on wavenumber
in $f(R)$ theories, $R_{ij}$ are the linear theory response functions,
and $\alpha({\bf k}_1,{\bf k}_2)$ and $\beta({\bf k}_1,{\bf k}_2)$ are the usual geometric
kernels that arise from the continuity and Euler equations.
The factor $\gamma^s_{2;11}$ is a new cubic vertex that arises from the fact that
in modified-gravity theories the effective gravitational potential becomes
a nonlinear functional of the matter density field (because of the contribution of the
fifth force), and $\gamma^s_{2;11}$ gives the quadratic contribution.
Thus, we define the spherically symmetric coefficient
\bea
&& \nu_2(k_1,k_2) = \frac{1}{D_+(k_1,\eta) D_+(k_2,\eta)}
\int_{-\infty}^{\eta} d\eta' \int_{-1}^{1} \frac{d\mu}{2} \biggl [
R_{11}(k;\eta,\eta') \biggl ( \alpha({\bf k}_1,{\bf k}_2)
\frac{\partial D_+}{\partial\eta}(k_1,\eta') D_+(k_2,\eta')
\nonumber \\
&& + \alpha({\bf k}_2,{\bf k}_1)  \frac{\partial D_+}{\partial\eta}(k_2,\eta')
D_+(k_1,\eta') \biggl ) + 2 R_{12}(k;\eta,\eta') \biggl ( \beta({\bf k}_1,{\bf k}_2)
\frac{\partial D_+}{\partial\eta}(k_1,\eta') \frac{\partial D_+}{\partial\eta}(k_2,\eta')
\nonumber \\
&& + \gamma^s_{2;11}({\bf k}_1,{\bf k}_2;\eta') D_+(k_1,\eta') D_+(k_2,\eta')
\biggl ) \biggl ] ,
\label{eq:nu2}
\eea
where ${\bf k}={\bf k}_1+{\bf k}_2$ and $\mu=({\bf k}_1\cdot{\bf k}_2)/(k_1 k_2)$.
For $k_1 \gg k_2$ this gives the amplification by long modes $k_2$ of the small-scale
modes $k\simeq k_1$, which provides the bias at lowest order of perturbation theory as in
Eq.(\ref{eq:nu2-def})  \cite{Seljak_2012}.
For $k_1=k_2$ this gives the leading-order contribution to the skewness
$\langle \delta^3 \rangle_c / \langle \delta^2\rangle^2_c$ at scale $k_1$.
For cosmologies that are scale independent, like LCDM and the K-mouflage model in the unscreened
regime, $\nu_2$ is independent of scale. For cosmologies that introduce new scales,
such as the $f(R)$ theories, it depends on scale.
In the case of the Einstein-de Sitter universe, we recover the usual coefficient
$\nu_2=34/21$ \cite{Bernardeau2002}.

\section{The Zeldovich approximation}
\label{app:Zeldovich}

In the Zeldovich approximation \cite{ZelDovich1970},
particles initially at a position ${\bf{q}}$ evolve along a trajectory
${\bf {x}}(t,{\bf q})$. The variances of the relative displacement between two initial
points separated by $\Delta {\bf q}$ are given by
\be
\sigma^2_\parallel (\Delta q)= 2 \int d{\bf k} \; [ 1- \cos(k_\parallel \Delta q) ]
\frac{k^2_\parallel}{{\bf k}^4} P_L(k)
\ee
along the initial separation, and
\be
\sigma^2_\perp (\Delta q)= 2 \int d{\bf k} \; [1- \cos(k_\perp \Delta q) ]
\frac{k^2_\perp}{{\bf k}^4} P_L(k)
\ee
in the orthogonal direction.
The linear power spectrum is given by $P_L(k)$.
In the Zeldovich approximation, and for Gaussian initial conditions, this leads to the Zeldovich
power spectrum \cite{Schneider1995,Taylor1996}
\be
P_{\rm Z}(k) = \int \frac{d{\bf \Delta q}}{(2\pi)^3} \;
e^{{\rm i} k \mu \Delta q - \frac{1}{2} k^2 \mu^2 \sigma^2_\parallel
- \frac{1}{2} k^2 (1-\mu^2) \sigma^2_\perp} ,
\ee
where the directing cosine is defined as $\mu = ( {\bf k} \cdot \Delta {\bf q} ) / (k \Delta q)$.
In the main text we use a truncated Zeldovich power spectrum (\ref{eq:PL-trunc}).

\bibliography{ref1}   

\end{document}